\documentclass[12pt, letterpaper]{JHEP3}

\usepackage{amssymb, amsmath, amsopn, amsthm}

\usepackage{epsfig}

\usepackage{epsfig,multicol}

\usepackage{amsmath}
\usepackage{graphicx}
\usepackage{latexsym}
\usepackage{amsfonts}
\usepackage{amssymb}

\newcommand{\eref}[1]{(\ref{#1})}

\newcommand{\fref}[1]{Figure~\ref{#1}}
\newcommand{\cref}[1]{Chapter~\ref{#1}}
\newcommand{\beq}{\begin{equation}}
\newcommand{\eeq}{\end{equation}}
\newcommand{\ba}{\begin{array}}
\newcommand{\ea}{\end{array}}
\newcommand{\bcenter}{\begin{center}}
\newcommand{\ecenter}{\end{center}}

\def\IB{\relax\hbox{$\inbar\kern-.3em{\rm B}$}}
\def\IC{\relax\hbox{$\inbar\kern-.3em{\rm C}$}}
\def\ID{\relax\hbox{$\inbar\kern-.3em{\rm D}$}}
\def\IE{\relax\hbox{$\inbar\kern-.3em{\rm E}$}}
\def\IF{\relax\hbox{$\inbar\kern-.3em{\rm F}$}}
\def\IG{\relax\hbox{$\inbar\kern-.3em{\rm G}$}}
\def\IGa{\relax\hbox{${\rm I}\kern-.18em\Gamma$}}
\def\IH{\relax{\rm I\kern-.18em H}}
\def\IK{\relax{\rm I\kern-.18em K}}
\def\IL{\relax{\rm I\kern-.18em L}}
\def\IP{\relax{\rm I\kern-.18em P}}
\def\IR{\relax{\rm I\kern-.18em R}}
\def\IZ{\relax\ifmmode\mathchoice
{\hbox{\cmss Z\kern-.4em Z}}{\hbox{\cmss Z\kern-.4em Z}}
{\lower.9pt\hbox{\cmsss Z\kern-.4em Z}}
{\lower1.2pt\hbox{\cmsss Z\kern-.4em Z}}\else{\cmss Z\kern-.4em Z}\fi}
\def\II{\relax{\rm I\kern-.18em I}}

\def\sCC{{\kern 0.27em\vrule height1.45ex width0.03em depth0em
          \kern-0.30em\rm C}}
\def\C{{\mathchoice
  {\sCC}
  {\sCC}
  {\kern 0.225em \vrule height1.05ex width0.025em depth0em \kern-0.25em \rm C}
  {\kern 0.180em \vrule height0.78ex width0.02em depth0em \kern-0.2em \rm C}
        }}
\def\sHH{{\rm I\kern-.16em{}H}}
\def\H{{\mathchoice
  {\sHH}
  {\sHH}
  {\rm I\kern-.13em{}H}
  {\rm I\kern-.13em{}H} }}
\def\sNN{{\rm I\kern-.16em{}N}}
\def\N{{\mathchoice
  {\sNN}
  {\sNN}
  {\rm I\kern-.12em{}N}
  {\rm I\kern-.10em{}N} }}
\def\sPP{{\rm I\kern-.16em{}P}}
\def\P{{\mathchoice
  {\sPP}
  {\sPP}
  {\rm I\kern-.12em{}P}
  {\rm I\kern-.10em{}P} }}
\def\sQQ{{\kern 0.27em \vrule height1.45ex width0.03em depth0em
          \kern-0.30em \rm Q}}
\def\Q{{\mathchoice
        {\sQQ}
        {\sQQ}
  {\kern 0.225em \vrule height1.05ex width0.025em depth0em \kern-0.25em \rm Q}
  {\kern 0.180em \vrule height0.78ex width0.020em depth0em \kern-0.20em \rm Q}
        }}
\def\sRR{{\rm I\kern-0.16em{}R}}
\def\R{{\mathchoice
  {\sRR}
  {\sRR}
  {\rm I\kern-0.12em{}R}
  {\rm I\kern-0.10em{}R} }}
\def\sZZ{{\rm Z\kern-0.32em{}Z}}
\def\Z{{\mathchoice
  {\sZZ}
  {\sZZ} 
  {\rm Z\kern-0.3em{}Z}     
  {\rm Z\kern-0.25em{}Z} }}  
\def\ZZZ{{\rm Z\kern-0.24em{}Z}}
\def\sII{{\rm I\kern-0.16em{}I}}
\def\I{{\mathchoice
  {\sII}
  {\sII}
  {\rm I\kern-0.12em{}I}
  {\rm I\kern-0.10em{}I} }}

\def\inbar{\,\vrule height1.5ex width.4pt depth0pt}
\font\cmss=cmss10 \font\cmsss=cmss10 at 7pt

\def\smiley{\hbox{\large$\bigcirc$\hspace{-0.80em}\raise.2ex
\hbox{$\cdot\cdot$}\kern-.61em\lower.2ex\hbox{\scriptsize$\smile$}}\ }
\def\frowny{\hbox{\large$\bigcirc$\hspace{-0.80em}\raise.2ex
\hbox{$\cdot\cdot$}\kern-.635em\lower.2ex\hbox{\scriptsize$\frown$}}\ }

\def\I{{\rlap{1} \hskip 1.6pt \hbox{1}}}

\makeatletter
\let\hangafter\@hangfrom
\makeatother

\def\makeatletter{\catcode`\@=11}
\makeatletter
\def\mathbox#1{\hbox{$\m@th#1$}}%
\def\math@ccstyles#1#2#3#4#5#6#7{{\leavevmode
     \setbox0\mathbox{#6#7}%
     \setbox2\mathbox{#4#5}%
     \dimen@ #3%
     \baselineskip\z@\lineskiplimit#1\lineskip\z@
     \vbox{\ialign{##\crcr
            \hfil \kern #2\box2 \hfil\crcr
            \noalign{\kern\dimen@}%
            \hfil\box0\hfil\crcr}}}}
\def\mathaccstyles{\math@ccstyles\maxdimen}
\def\maththroughstyles{\math@ccstyles{-\maxdimen}}
\def\unity%
{\maththroughstyles{.45\ht0}\z@\displaystyle {\mathchar"006C}\displaystyle 1}

\renewcommand{\H}{\mathcal{H}}

\newcommand{\be}{\begin{eqnarray}}
\newcommand{\bea}{\begin{eqnarray}}
\newcommand{\ee}{\end{eqnarray}}
\newcommand{\eea}{\end{eqnarray}}

\newcommand{\bb}{\mathsf{b}}
\newcommand{\ww}{\mathsf{w}}

\title{Dimer Models, Integrable Systems and Quantum Teichm\"uller Space}

\author{Sebasti\'an Franco

\\

\vspace{0.2cm}
~\\

Kavli Institute for Theoretical Physics University of California \\
Santa Barbara, CA 93106, USA \\
\vspace{0.2cm}

\email{sfranco@kitp.ucsb.edu}\\

}

\abstract{We introduce a correspondence between dimer models (and hence superconformal quivers) and the quantum Teichm\"uller space of the Riemann surfaces associated to them by mirror symmetry. Via the untwisting map, every brane tiling gives rise to a tiling of the Riemann surface with faces surrounding punctures. We explain how to obtain an ideal triangulation by dualizing this tiling. In order to do so, tiling nodes of valence greater than 3 (equivalently superpotential terms of order greater than 3 in the corresponding quiver gauge theories) must be decomposed by the introduction of 2-valent nodes. From a quiver gauge theory perspective, this operation corresponds to integrating-in massive fields. Fock coordinates in Teichm\"uller space are in one-to-one correspondence with chiral fields in the quiver. We present multiple explicit examples, including infinite families of theories, illustrating how the right number of Fock coordinates is generated by this procedure. Finally, we explain how Chekhov and Fock commutation relations between coordinates give rise to the commutators associated to dimer models by Goncharov and Kenyon in the context of quantum integrable systems. For generic dimer models (i.e. those containing nodes that are not 3-valent), this matching requires the introduction of a natural generalization of Chekhov and Fock rules. We also explain how urban renewal in the original brane tiling (Seiberg duality for the quivers) is mapped to flips of the ideal triangulation.
}

\preprint{NSF-KITP-11-068}

\begin{document}

\tableofcontents

\section{Introduction}

The study of $d=4$, $\mathcal{N}=1$, superconformal quiver gauge theories arising on D3-branes probing toric Calabi-Yau singularities has been immensely simplified by the discovery of a correspondence connecting them to dimer models \cite{Franco:2005rj}. 

Over the years, this correspondence has ramified in multiple directions, densely covering the gap between mathematics and physics. To give a flavor of its diverse applications, we can mention: computation of Donaldson-Thomas invariants for general toric geometries via crystal melting \cite{Ooguri:2008yb}, mirror symmetry \cite{Feng:2005gw}, toric/Seiberg duality \cite{Franco:2005rj}, non-perturbative effects in string theory (D-brane instantons) \cite{Franco:2007ii}, SUSY breaking in string theory \cite{Franco:2007ii}, AdS/CFT correspondence \cite{Franco:2005sm,Butti:2005sw} and local embeddings of the MSSM and flavor physics in string theory \cite{Krippendorf:2010hj}.

One of the reasons that make dimers models so outstanding is that they not only define an infinite set of interesting objects (the largest known classification of 4d superconformal gauge theories) but also make some previously complicated calculation (the computation of their moduli space) trivial. One can certainly wonder whether anything similar can be achieved once again. The answer is yes. Remarkably, Goncharov and Kenyon have shown that dimer models define an infinite set of 0+1 dimensional quantum integrable systems \cite{GK}. In addition, constructing all their conserved charges is straightforward using dimer models.

In this paper we will focus on one specific byproduct of the new correspondence. As originally pointed out by Goncharov and Kenyon \cite{GK}, there is a profound similarity between dimer models and the Teichm\"uller space of Riemann surfaces. Furthermore, the new correspondence hints to connections to quantum Teichm\"uller theory. The purpose of this paper is to work out the details of this connection.

This work is organized as follows. Section 2 summarizes many of the ingredients that are used in the paper, including quiver gauge theories, dimer models, integrable systems and mirror symmetry. Section 3 reviews Teichm\"uller space and its quantization. Section 4 explains how to get ideal triangulations of Riemann surfaces and define coordinates on Teichm\"uller space using dimer models. Section 5 connects the Chekhov-Fock quantization of Teichm\"uller space to the commutation relations introduced by Goncharov and Kenyon in the context of the correspondence between dimer models and quantum integrable systems. We also explain how flips in the ideal triangulation arise from urban renewal in the brane tiling (Seiberg duality). Section 6 shows how Goncharov-Kenyon commutators follow from Chekhov-Fock rules in various explicit examples. We conclude in Section 7. For reference, we include an appendix summarizing the geometry, gauge theory and brane tiling for each of the explicit examples considered in the paper.

\section{Cast of Characters}

\subsection{Dimer Models, Quiver Gauge Theories and Toric Singularities}

D3-branes probing toric Calabi-Yau singularities give rise to $d=4$, $\mathcal{N}=1$ quiver gauge theories on their worldvolume. A correspondence between these gauge theories and dimer models was introduced in \cite{Franco:2005rj}. Dimer models are bipartite graphs living on a $\mathbb{T}^2$. In string theory, they correspond to physical configurations of NS5 and D5-branes, twice T-dual to the D3-branes on the singularity. For this reason we also refer to them as {\it brane tilings}. In what follows, we refer to the brane tiling as $T$.

Brane tilings combine quiver and superpotential information into a single object. The dictionary of the correspondence reads:

\bigskip

\beq
\begin{array}{ccccc}
\mbox{{\bf Gauge Theory}} & & & & \mbox{{\bf Brane Tiling}} \\
\mbox{gauge group} & \ \ \ \ & \leftrightarrow & \ \ \ \ & \mbox{face}\\
\mbox{chiral superfield} & \ \ \ \ & \leftrightarrow & \ \ \ \ & \mbox{edge} \\
\mbox{superpotential term} & \ \ \ \ & \leftrightarrow & \ \ \ \ & \mbox{node}
\end{array}
\nonumber
\eeq

\bigskip

The two independent cycles of the 2-torus correspond to the $U(1)^2$ flavor symmetry common to all these theories, which follows from the isommetries of the underlying toric Calabi-Yaus. 

Gauged linear sigma model fields in the toric construction of the moduli space of the gauge theories are in one to one correspondence with perfect matchings of the dimer model. Their position in the toric diagram is given by the slope of the height function. As a result of this mapping, the computation of the moduli space is greatly simplified, and is reduced to taking the determinant of the Kasteleyn matrix. Thanks to this simplification, dimer models have played an instrumental role in the determination of infinite families of explicit AdS/CFT dual pairs \cite{Franco:2005sm,Butti:2005sw}.

The correspondence connecting dimer model, quiver theories and toric singularities has been explained using mirror symmetry in \cite{Feng:2005gw} and proved in \cite{Franco:2006gc}.

\subsection{Dimer Models and Integrable Systems}

\label{section_integrable_dimers}

We now provide a brief summary of the correspondence between dimer models and integrable systems introduced in \cite{GK}. For clarity, we illustrate the ideas in the explicit case of the phase II of $F_0$. Details of this theory can be found in the appendix.

\newpage

\begin{center} {\bf 1 - Dynamical Variables} \end{center}

\smallskip

The variables of the problem correspond to closed cycles on the brane tiling. A convenient basis for cycles is:
\medskip
\beq
\begin{array}{ccc}
\mbox{{\bf \underline {Basis of cycles}:}} & \ \ \ & 
\bullet \ \mbox{$w_i$ ($i=1,\ldots, N_g$): cycles going clockwise around each face.} \\ 
 & & \bullet \ \mbox{$z_1$ and $z_2$: cycles winding around the two torus directions.}
\end{array}
\nonumber
\eeq
\medskip
Here $N_g$ is the number of gauge groups in the quiver, i.e. the number of faces in the tiling.\footnote{Since $\prod_{i=1}^G w_i=1$, one of the $w_i$'s is actually redundant.} \fref{cycles_F0_II} shows these cycles for phase II of $F_0$.  

\begin{figure}[h]
\begin{center}
\includegraphics[width=11cm]{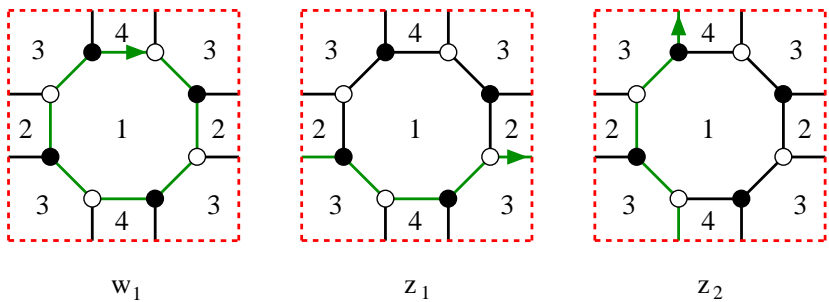}
\caption{Some of the basic cycles for phase II of $F_0$.}
\label{cycles_F0_II}
\end{center}
\end{figure}

It is important to emphasize that while the $w_i$ and $z_j$ variables provide a natural basis of cycles on the tiling, any other basis is equally valid. In some cases, a judicious choice of basis nicely simplifies the problem.

\smallskip

\begin{center} {\bf 2 - Poisson Structure} \end{center}

\smallskip

\begin{figure}[ht]
\begin{minipage}[h]{0.5\linewidth}
\centering
\includegraphics[scale=.57]{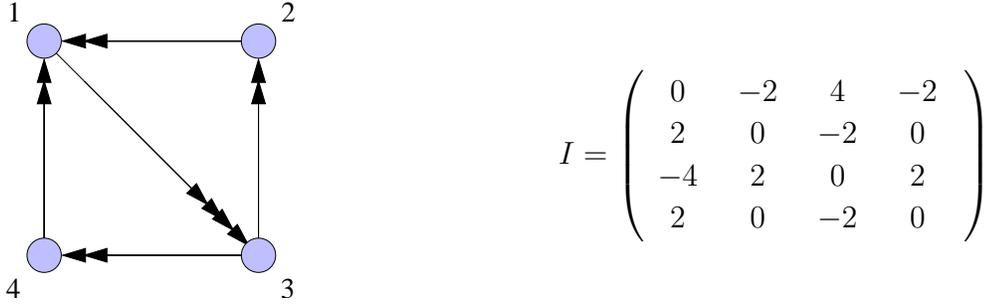}
\label{fig:figure1}
\end{minipage}
\hspace{0.25cm}
\begin{minipage}[h]{0.5\linewidth}
\centering 
\beq
I=\left(\begin{array}{cccc} \ \ 0 \ \ & \ \ -2 \ \ & \ \ 4 \ \ & \ \ -2 \ \ \\
2 & 0 & -2 & 0 \\
-4 & 2 & 0 & 2 \\
2 & 0 & -2 & 0 \end{array}\right)\nonumber
\eeq
\end{minipage}
\caption{Quiver diagram and adjacency matrix for phase II of $F_0$.}
\label{quiver_F0}
\end{figure}

The next step consists of defining a Poisson structure. We have

\beq
\{w_i,w_j\}=\epsilon_{ij} \, w_i w_j \, .
\eeq
In this expression and the ones that follow, $\epsilon_{ij}$ counts the number of edges over which two paths overlap, weighted by their orientation. This implies that we can write
\beq
\{w_i,w_j\}=I_{ij} \, w_i w_j \, ,
\label{PB_ww}
\eeq
where $I_{ij}$ is the antisymmetric oriented incidence matrix of the quiver. \fref{quiver_F0} shows the quiver and incidence matrix for our example. For example, since nodes 1 and 3 are connected by four bifundamental fields going out of node 4, we have $\{w_1,w_3\}=4 \, w_1 w_3$. Similarly, nodes 1 and 2 are connected by two bifundamentals coming into node 1, which results in $\{w_1,w_2\}=-2 \, w_1 w_2$.

The cycles associated with $z_1$ and $z_2$ intersect an odd number of times, which give an extra contribution to their Poisson bracket, in addition to the edge overlap. 

\beq
\{z_1,z_2\}=1+\epsilon_{z_1,z_2} \, .
\label{PB_zz}
\eeq
Finally, we have
\beq
\{z_a,w_i\}=\epsilon_{z_a,w_i} \,.
\label{PB_zw}
\eeq

\begin{figure}[h]
\begin{center}
\includegraphics[width=12cm]{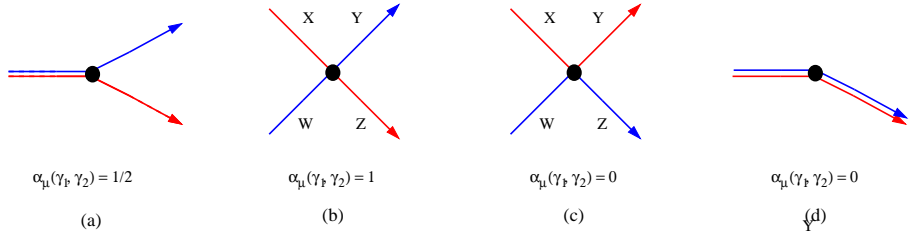}
\caption{Local contributions to Poisson brackets. We show $\gamma_1$ and $\gamma_2$ in red and blue, respectively.}
\label{local_rules_commutators}
\end{center}
\end{figure}

Alternatively, Poisson brackets can also be encoded in terms of local rules \cite{GK}. Denoting $x_i$ the variable associated to a path $\gamma_i$, we have

\beq
\{x_1,x_2\}= \alpha_{1,2} \, x_{\gamma_1} x_{\gamma_2} \, ,
\eeq
where 
\beq
\alpha_{1,2}=\sum_\mu \alpha_\mu(\gamma_1,\gamma_2)
\eeq
is a sum of the contributions summarized in \fref{local_rules_commutators} over all nodes indexed by $\mu$.

The sign of each contribution is inverted whenever the direction of an arrow is reversed or the color of the node is changed. Notice that while \fref{local_rules_commutators} only shows edges belonging to $\gamma_1$ and $\gamma_2$, there might be additional edges terminating on the nodes under consideration (i.e. we do not restrict to tilings containing only 3 and 4-valent nodes). 

The quantum theory is nicely encoded in terms of a q-deformed algebra. Defining $X_i=e^{x_i}$, we get
\beq
X_i X_{j}=q^{\alpha_{ij}} X_j X_{i} \, ,
\label{quantum_commutators_GK}
\eeq
where $q=e^{-i2\pi\hbar}$ and we have promoted Poisson brackets to commutators in the usual way: $[A,B]=i \, 2 \pi \, \hbar \{A,B\}$.\footnote{To keep the notation simple, we do not distinguish quantum operators from their corresponding classical variables throughout the paper. Whether we are discussing a classical variable or a quantum operator should be clear from the context.}

\smallskip

\begin{center} {\bf 3 - Conserved Charges} \end{center}

\smallskip

Every perfect matching is associated to a point in the toric diagram and defines a closed loop by taking its difference with a reference perfect matching. The resulting loop can be expressed as a product of appropriate powers of the basic loops $\{w_i,z_1,z_2\}$. When more than one perfect matching correspond to the same point in the toric diagram their contributions must be added. Very briefly, Goncharov and Kenyon \cite{GK} have proved that, with the commutation relations \eref{quantum_commutators_GK}, this procedure results in:

\begin{itemize}
\item {\bf \underline{Casimirs}:} they correspond to ratios of consecutive points on the boundary of the toric diagram. They commute with everything.

\item {\bf \underline{Hamiltonians}:} they correspond to internal points in the toric diagram. They commute with each other.
\end{itemize}

Consider a toric diagram with $n_{interior}$ internal points. Using the conserved Casimirs to eliminate variables, we are left with a $2\, n_{interior}$-dimensional phase space. Since the number of Hamiltonians is $n_{interior}$, we conclude the construction defines a $(0+1)$-dimensional {\it quantum integrable system}!

Every coefficient in the characteristic polynomial $P(z_1,z_2)$ can be associated to a conserved charge. The Riemann surface defined by the equation $P(z_1,z_2)=0$, which we denote $\Sigma$, is the {\it spectral curve} of the integrable system. The main objective of this paper is to investigate the Teichm\"uller space of $\Sigma$. In the coming section we review the important role played by $\Sigma$ in the mirror of the original singularity. 

\bigskip

We will not explicitly pursue the fascinating connection between dimer models and integrable systems in this paper. A thorough study of this correspondence from a physics perspective will appear in a forthcoming publication \cite{Mina_et_al}. The commutation relations \eref{quantum_commutators_GK} will play a central role when connecting dimer models to the quantization of Teichm\"uller space in Section \ref{section_QT_from_dimers}.

\bigskip

\subsection{Mirror Symmetry, Riemann Surfaces and the Untwisting Map}

\label{section_mirror}

Consider a toric singularity with characteristic polynomial $P(z_1,z_2)=\sum a_{n1,n2} \, z_1^{n_1}z_2^{n_2}$, where $(n_1, n_2)$ runs over points in the toric diagram.
The mirror manifold is given by $P(z_1,z_2)=W$, $W= u v$. The spectral curve $\Sigma$ of the associated integrable system is the Riemann surface sitting at 
$W=0$ and it plays a crucial role in the derivation of dimer models using mirror symmetry \cite{Feng:2005gw}. The genus and number of punctures of $\Sigma$ are equal to the number of internal points and perimeter of the toric diagram, respectively. 

\begin{figure}[h]
\begin{center}
\includegraphics[width=4cm]{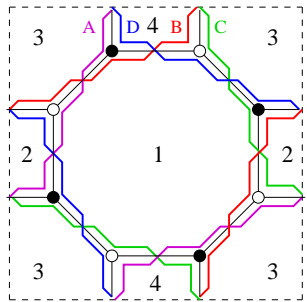}
\caption{Double line implementation of zig-zag paths for phase II of $F_0$.}
\label{double_line_F0_II}
\end{center}
\end{figure}

Among all possible paths on a brane tiling, a prominent role is played by the so called {\it zig-zag} paths, which alternatively turn maximally right and left at each node. Zig-zag paths are beautifully implemented using a double line notation for edges \cite{Feng:2005gw}. \fref{double_line_F0_II} shows the double line implementation of zig-zag paths for phase II of $F_0$. The double line notation naturally singles out the middle point of edges, something that will reappear later in the paper.

\begin{figure}[h]
\begin{center}
\includegraphics[width=8cm]{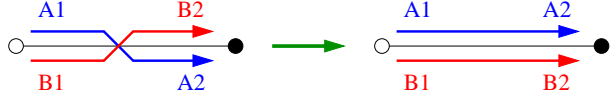}
\caption{The untwisting map.}
\label{untwisting}
\end{center}
\end{figure}

Consider the {\it untwisting map} \cite{Feng:2005gw}, whose action is schematically shown in \fref{untwisting}. The map transforms the brane tiling $T$ into another bipartite graph $\tilde{T}$ that tiles the spectral curve $\Sigma$.\footnote{Notice that generically, neither $\Sigma$ is a $\mathbb{T}^2$ (with punctures) nor $\tilde{T}$ is the same kind of lattice as the original $T$.} Applying the untwisting map to the zig-zag paths of $\tilde{T}$ takes us back to $T$. Its action interchanges:

\beq
\begin{array}{ccccc}
\mbox{$T$ on 2-torus} & & & & \mbox{$\tilde{T}$ on $\Sigma$} \\
\mbox{zig-zag path} & \ \ \ \ & \leftrightarrow & \ \ \ \ & \mbox{face (puncture)}\\
\mbox{face (gauge group)} & \ \ \ \ & \leftrightarrow & \ \ \ \ & \mbox{zig-zag path}
\end{array}
\nonumber
\eeq

\bigskip

\begin{figure}[h]
\begin{center}
\includegraphics[width=8.5cm]{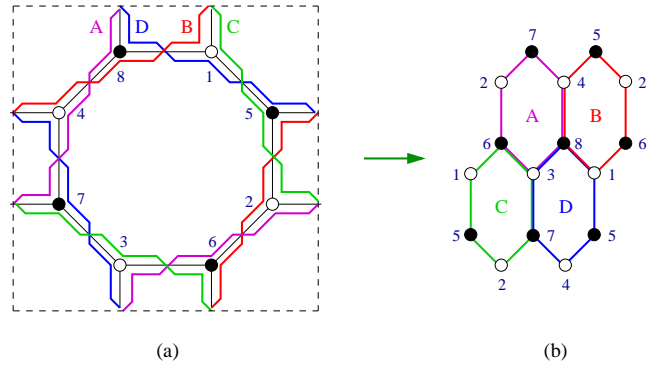}
\caption{a) Brane tiling for phase II of $F_0$, with the four zig-zag paths indicated in double line notation. b) After untwisting, we obtain a hexagonal tiling $\tilde{T}$ of $\Sigma$, which in this case is a 2-torus with four punctures ($A$, $B$, $C$ and $D$).}
\label{untwisting_F0_II}
\end{center}
\end{figure}

Let us illustrate how the untwisting map acts on \fref{double_line_F0_II}. The result is shown in \fref{untwisting_F0_II}. In this case, $\Sigma$ is a 2-torus with four punctures (that we label $A$, $B$, $C$ and $D$) and $\tilde{T}$ is a hexagonal lattice. Notice that while both the original tiling $T$ and $\tilde{T}$ have 3-valent nodes, they are completely different lattices. This example was first studied in detail in \cite{Feng:2005gw}. Sections \ref{section_QT_from_dimers} and \ref{section_examples} present several additional examples.

\begin{figure}[h]
\begin{center}
\includegraphics[width=9.5cm]{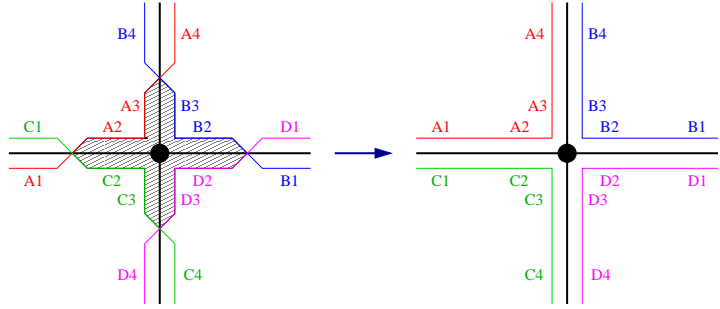}
\caption{Effect of the untwisting map around a node. The area surrounding the node (shaded in the figure) is invariant under the map. While we show the specific case of a quartic node, this behavior is completly general.}
\label{untwisting_node}
\end{center}
\end{figure}

A general property of the untwisting map is that it preserves the local structure around a node. Everything at a distance smaller than half an edge from a given node remains invariant. This implies that not only the fields ending on a node (and as a result its valence) but also their cyclic ordering around it are preserved. \fref{untwisting_node} shows the example of a 4-valent node. Also, close paths on $T$ map to closed paths on $\tilde{T}$.

\subsection{From Edges to Paths on the Tiling}

\label{section_integrals_closed_paths}

In this section we describe the map between edges and closed paths on the tiling following \cite{Kenyon:2003uj}. Our discussion applies to both $T$ and $\tilde{T}$ without changes. The brane tiling is bipartite, giving each edge a natural orientation from its white vertex to its black vertex. Any function $\epsilon(e)$ on the edges defines a 1--form, satisfying $\epsilon(-e)=-\epsilon(e)$, where $-e$ indicates the edge with opposite direction.

Consider a length-$k$ closed path on the tiling
\beq
\gamma=\{\ww_0,\bb_0,\ww_1,\bb_1,\ldots,\bb_{k-1},\ww_k\} \ \ \ \ \ \ \ww_k=\ww_0 \, ,
\eeq
where $\ww_i$ and $\bb_i$ indicate white and black nodes, respectively. The magnetic flux through $\gamma$ is defined as
\beq
B(\gamma)=\int_\gamma \epsilon=\sum_{i=1}^{k-1}\left[\epsilon(\ww_i,\bb_i)- \epsilon(\ww_{i+1},\bb_i)\right] \, .
\eeq 
It is possible to define gauge transformation on the brane tiling (which should not to be confused with the quiver gauge symmetries). Magnetic fluxes are invariant under gauge transformation. The gauge inequivalent classes of 1--forms can be parametrized by the magnetic fluxes through $\gamma_{w_i}$, $\gamma_{z_1}$ and $\gamma_{z_2}$.\footnote{As explained in Section \ref{section_integrable_dimers}, one $\gamma_{w_i}$ is redundant.} These gauge transformations were exploited in \cite{Franco:2006gc} for solving F--term equations. From \cite{Franco:2006gc}, it is natural to relate complex 1--form to fields in the quiver in the following way
\beq
\epsilon(e_i) = \ln X_i \, .
\eeq

We can define new variables associated to closed paths, given by the exponentials of magnetic fluxes:
\beq
v(\gamma)=e^{\int_\gamma \epsilon}=\prod_{i=1}^{k-1} {X(\ww_i,\bb_i)\over X(\ww_{i+1},\bb_i)} \, ,
\label{flux_v}
\eeq
where the product runs over the contour $\gamma$. The loop variables in Section \ref{section_integrable_dimers} are indeed $w_j \equiv v(\gamma_{w_j})$, $z_1\equiv
v(\gamma_{z_1})$ and $z_2\equiv v(\gamma_{z_2})$. In addition, although not relevant for the discussion in this paper, the Hamiltonians are 
sum of $v(\gamma)$ variables for appropriate $\gamma$'s and the Casimirs are ratios of them. 

\section{Teichm\"uller Space}

Consider a Riemann surface $\Sigma_{g,n}$ with genus $g$ and $n$ punctures.
Its Teichm\"uller space is the space of complex structure deformations of $\Sigma_{g,n}$ divided by the identity component of the diffeomorphisms of $\Sigma_{g,n}$

\beq
\mathcal{T}_{g,n}={\mbox{Complex Structure on } \Sigma_{g,n} \over \mbox{Diff}_0(\Sigma_{g,n} )} \, .
\eeq

In order to define coordinates in Teichm\"uller space, we consider an {\it ideal triangulation} of $\Sigma_{g,n}$, namely a triangulation whose vertices are located at the punctures. Let us define $m$ to be minus the Euler characteristic, i.e. $m=-\chi(\Sigma_{g,n})=2g-2+n$.\footnote{The number of gauge groups of the corresponding quiver theory is equal to minus the Euler characteristic: $N_g=m=-\chi(\Sigma_{g,n})$.} The numbers of triangular faces $(F)$, edges $(E)$ and vertices $(V)$ in an ideal triangulation are:

\beq
F=2 \, m \ \ \ \ \ \ E=3 \, m \ \ \ \ \ \ \ V=n
\eeq
To each edge $e$ we associate a real Fock coordinate $z_e$ (also denoted {\it shear coordinate} ). Fock coordinates parametrize how ideal triangles are glued together to reconstruct the Riemann surface \cite{Fock}.

\begin{figure}[h]
\begin{center}
\includegraphics[width=8cm]{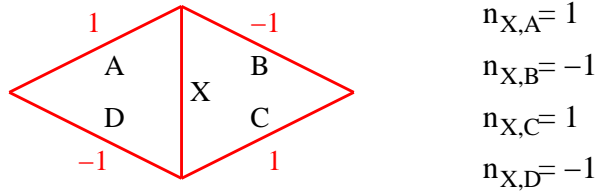}
\caption{This rhombus diagram summarizes the contributions to $n_{e,e'}$, which define the Poisson brackets between coordinates in Teichm\"uller space given by the Weil-–Petersson Poisson structure. The Poisson brackets are promoted to commutators in the quantum theory.}
\label{rhombus_Fock_rules}
\end{center}
\end{figure}

The Weil-–Petersson Poisson structure on Teichm\"uller space is given by the brackets

\beq
\{x_e,x_{e'}\}=n_{e,e'} \, ,
\eeq
where $n_{e,e'} \in \{-2,-1,0,1,2\}$ and is determined by summing the contributions summarized in \fref{rhombus_Fock_rules}. For example, if the periodicity is such that $A$ and $D$ are identified, then we have $n_{X,A\equiv D}=2$. We will refer to this prescription as Chekhov-Fock (CF) rule.

The Checkhov-Fock (CF) quantization of Teichm\"uller space promotes the Weil-–Petersson Poisson brackets to commutators \cite{CF} 
\beq
[x_e,x_{e'}]=i 2 \, \pi \hbar \, n_{e,e'} \, .
\eeq
Defining $X_e=e^{x_e}$, we get
\beq
X_e X_{e'}=q^{n_{e,e'}} X_e' X_{e} \, ,
\label{exponential_commutation_Fock}
\eeq
where $q=e^{-i2\pi\hbar}$. This expression has clearly the same structure of Goncharov-Kenyon (GK) commutators \eref{quantum_commutators_GK}. We will later show how GK commutators follow from CF ones.

\section{Teichm\"uller Space from Dimer Models}

In this section we explain the connection between dimer models and Teichm\"uller space. In order to do so, we first explain how to construct an ideal triangulation of $\Sigma$ starting from an arbitrary brane tiling $T$. We focus on Fock coordinates, discussing Kashaev coordinates at the end.

\subsection{From Dimers to Ideal Triangulations: 3-valent Tilings}

By construction, faces of $\tilde {T}$ (which correspond to zig-zags in $T$) are centered around punctures of $\Sigma$. For the moment, let us focus on the case in which all nodes in $T$ are 3-valent, which implies that all nodes in $\tilde{T}$ are also 3-valent. We conclude the graph dual to $\tilde{T}$ provides an ideal triangulation of $\Sigma$. As usual, we understand dualization as the operation that maps (face, edge, node) $\to$ (node, transverse edge, face).

Edges in the triangulation are in one-to-one correspondence with edges in $\tilde{T}$, which are mapped to chiral fields in the corresponding quiver gauge theories. Thus, we can associate a Fock coordinate (which is a real number) to every chiral field. 

There is a profound similarity between $\tilde{T}$ and the (skeletons of) fat graphs used by Fock \cite{Fock}. In fact, both objects coincide if we restrict our attention to cubic tilings and bipartite fat graphs. Notice that bipartiteness is not a necessary condition for fat graphs. Fock's starting point in \cite{Fock} is an ideal triangulation, naturally leading to cubic fat graphs. Our approach and goal are slightly different, in fact we go in the opposite direction, starting from dimer models and making contact with ideal triangulations. This naturally takes us beyond cubic graphs. The interesting structures that follow from this generalization are the subject of coming sections.

\bigskip

\subsection*{Examples}

\begin{figure}[h]
\begin{center}
\includegraphics[width=7.5cm]{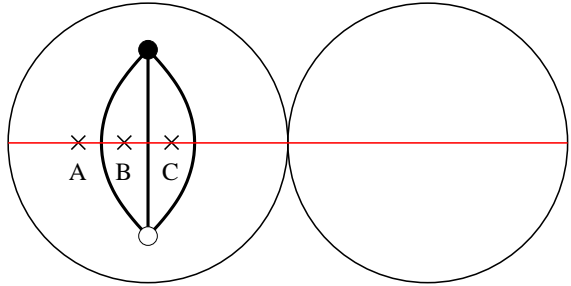}
\caption{The $\Sigma$ for $\mathbb{C}^3$ is a sphere with three punctures. The ideal triangulation obtained by dualizing $\tilde{T}$ is shown in red.}
\label{ideal_triangulation_C3}
\end{center}
\end{figure}

Let us look at some explicit examples. First, consider $\mathbb{C}^3$, whose corresponding gauge theory is $\mathcal{N}=4$ super Yang-Mills. This theory has three chiral fields and a superpotential consisting of two cubic terms. The associated Riemann surface is a sphere with three punctures. \fref{ideal_triangulation_C3} shows $\tilde{T}$ \cite{Feng:2005gw} and the ideal triangulation obtained by dualizing it. It consists of a single triangle dividing the sphere in two halves.

\begin{figure}[h]
\begin{center}
\includegraphics[width=4cm]{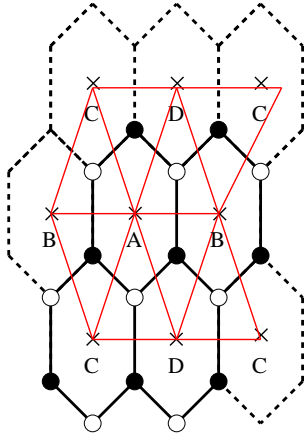}
\caption{The $\Sigma$ for $F_0$ is a 2-torus with four punctures. The ideal triangulation obtained by dualizing the $\tilde{T}$ for phase II of $F_0$ is shown in red.}
\label{ideal_triangulation_F0_II}
\end{center}
\end{figure}

Let us now consider phase II of $F_0$. It consists of twelve chiral fields and a purely cubic superpotential. As we have discussed in Section \ref{section_mirror} the Riemann surface for $F_0$ is a 2-torus with four punctures. In \fref{ideal_triangulation_F0_II} we present its $\tilde{T}$ \cite{Feng:2005gw} and the dual ideal triangulation. 

\subsection{From Dimers to Ideal Triangulations: the General Case}

The procedure outlined in the previous section requires modifications in order to deal with brane tilings containing k-valents nodes with $k>3$. Since the dual of a k-valent node is a k-sided polygon, the graph dual to $\tilde{T}$ is not a triangulation. 

\begin{figure}[h]
\begin{center}
\includegraphics[width=8.5cm]{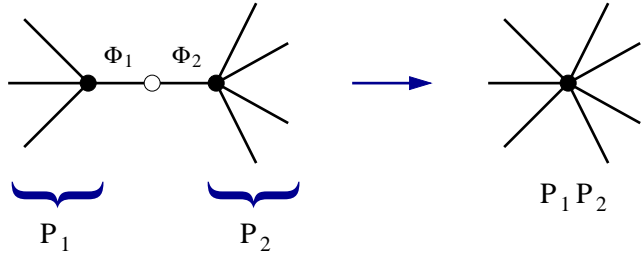}
\caption{A $(k_1+k_2-2)$-valent node is generated by collapsing a $k_1$ and a $k_2$-valent nodes separated by a 2-valent one. From a quiver perspective, this operation corresponds to integrating out $\Phi_1$ and $\Phi_2$.}
\label{integrating_out_2valent}
\end{center}
\end{figure}

This problem is solved by decomposing high valence nodes using 2-valent nodes. We refer the reader to \cite{Franco:2005rj} for a detailed discussion of this procedure. From a quiver perspective, a 2-valent node corresponds to a mass term for a pair of chiral fields $\Phi_1$ and $\Phi_2$ with opposite gauge charges. Let us discuss the inverse process, in which two nodes are merged when integrating out an intermediate 2-valent node. The corresponding configuration is shown in \fref{integrating_out_2valent}. The superpotential takes the form
\beq
W(\Phi_i)=\Phi_1 \Phi_2 - \Phi_1 P_1(\Phi_i)-\Phi_2 P_2 (\Phi_i)+\ldots \, ,
\eeq
where the dots indicate additional terms in the superpotential that do not involve $\Phi_1$ or $\Phi_2$ and $P_1$ and $P_2$ are products of fields that do not include $\Phi_1$ or $\Phi_2$. Removing a 2-valent node and the corresponding edges corresponds to integrating out the massive fields using F-term equations. As a result, we obtain
\beq
W(\Phi_i)=- P_1(\Phi_i) P_2 (\Phi_i)+\ldots \, .
\eeq
In terms of dimers, this means that two nodes of order $k_1$ and $k_2$ are combined into a new node of order $(k_1+k_2-2)$ as in \fref{integrating_out_2valent}. In what follows, we decompose nodes by reversing this process. 

\begin{figure}[h]
\begin{center}
\includegraphics[width=13cm]{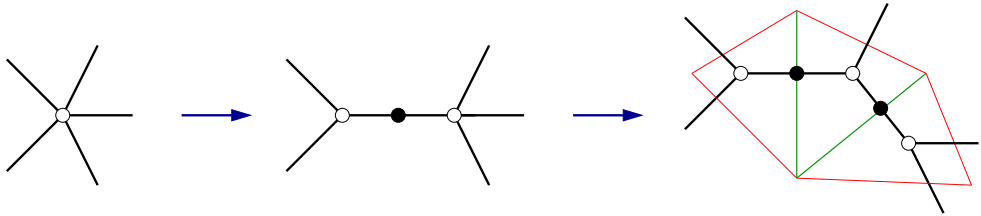}
\caption{One of the possible decompositions of a 5-valent node. In the last figure, we show the ideal triangles that we obtain, with double edges indicated in green.}
\label{5-valent_decomposition}
\end{center}
\end{figure}

Given a high valence node, there are multiple ways of decomposing it but all of them are equivalent. From a quiver gauge theory perspective this statement is trivial: the low energy physics that results from integrating out the massive fields associated to 2-valents nodes is unique.\footnote{Mathematically, it is interesting to mention that no new perfect matchings or zig-zag paths are generated by the addition of 2-valent nodes and the existing ones are trivially modified.} A k-valent node is decomposed into $(k-2)$ 3-valent and $(k-3)$ 2-valent ones. \fref{5-valent_decomposition} shows one of the possible decompositions of a 5-valent node.

\begin{figure}[h]
\begin{center}
\includegraphics[width=9cm]{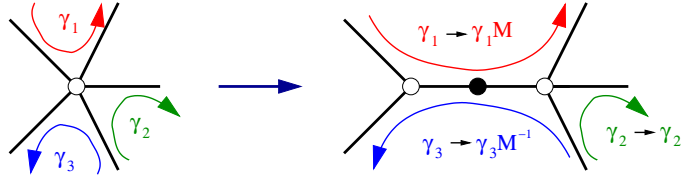}
\caption{Some examples of how paths get modified after introducing 2-valent vertices.}
\label{M_insertion}
\end{center}
\end{figure}

The last step is to associate a single edge, which we call a {\it double edge}, to every 2-valent node and its corresponding edges. The result of this construction is an ideal triangulation of $\Sigma$. \fref{5-valent_decomposition} shows how ideal triangles are generated by node decomposition. Paths in the tiling must be modified by appropriately inserting the edges $\Phi^{(\mu)}_1$ and $\Phi^{(\mu)}_2$, $\mu=1,\ldots,n_{2-valent}$, when necessary. Interestingly, the new edges only enter paths via the combination $M^{\mu}=\Phi^{(\mu)}_1/\Phi^{(\mu)}_2$. This fact justifies associating a single double edge to every 2-valent node. \fref{M_insertion} shows examples illustrating how paths transform.

We conclude that every double edge introduces a single coordinate in Teichm\"uller space. This implies that the number of Fock coordinates, which is equal to $3N_g$, can also be written as

\beq
3 N_g=N_f+\sum_{a\in W} (k_a-3) \, .
\label{dT_gauge_theory}
\eeq
We denote $N_g$, $N_f$ and $N_W$ the numbers of gauge groups, fields and superpotential terms in the quiver. This equation is not surprising. It follows from the facts that in toric quivers every field appears exactly twice in the superpotential, hence $\sum_{a\in W}k_a=2N_f$, and the relation $N_{f}=N_{g}+N_W$ holds \cite{Franco:2005rj}.\footnote{Equation \eref{dT_gauge_theory} can be regarded as a simple constraint on the structure (order of terms) of the superpotential of toric quivers obtained from integrating-in massive fields. We are not aware of this statement having appeared in the literature.}

\begin{figure}[h]
\begin{center}
\includegraphics[width=15cm]{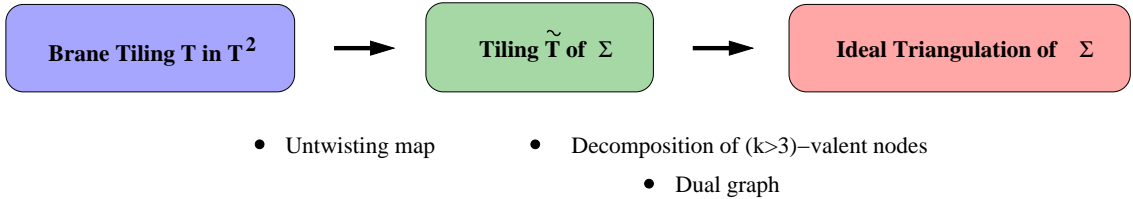}
\caption{Construction of an ideal triangulation starting for an arbitrary brane tiling.}
\label{flow_chart}
\end{center}
\end{figure}

The construction of an ideal triangulation starting for an arbitrary brane tiling can be summarized in the flow chart in \fref{flow_chart}.

\bigskip

\subsection*{Examples}

\begin{figure}[h]
\begin{center}
\includegraphics[width=15cm]{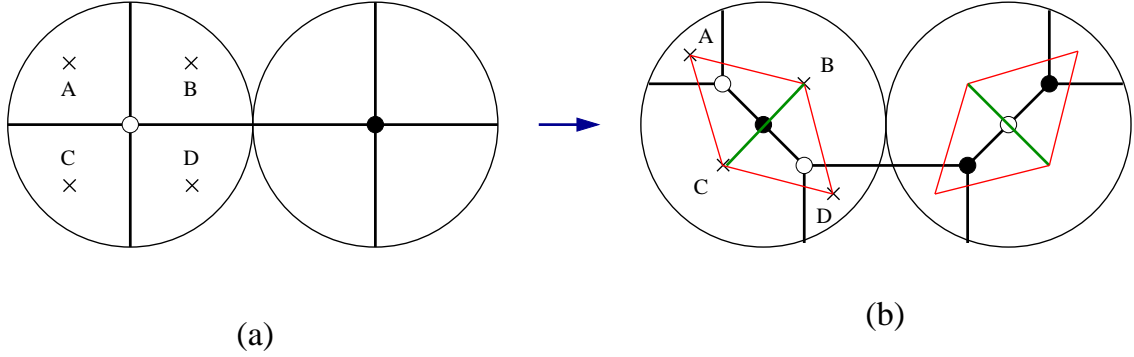}
\caption{The Riemann surface for the conifold is a sphere with four punctures. a) $\tilde{T}$ contains four edges and two quartic nodes. b) $\tilde{T}$ and dual triangulation after one possible decomposition of the quartic nodes. Double edges are indicated in green.}
\label{ideal_triangulation_conifold}
\end{center}
\end{figure}

The conifold gauge theory contains four fields and two quartic superpotential terms. The corresponding Riemann surface is a sphere with four punctures. \fref{ideal_triangulation_conifold} shows $\tilde{T}$ \cite{Feng:2005gw} and the ideal triangulation obtained by dualizing it after one of the possible decompositions of the quartic nodes. 

\begin{figure}[h]
\begin{center}
\includegraphics[width=11cm]{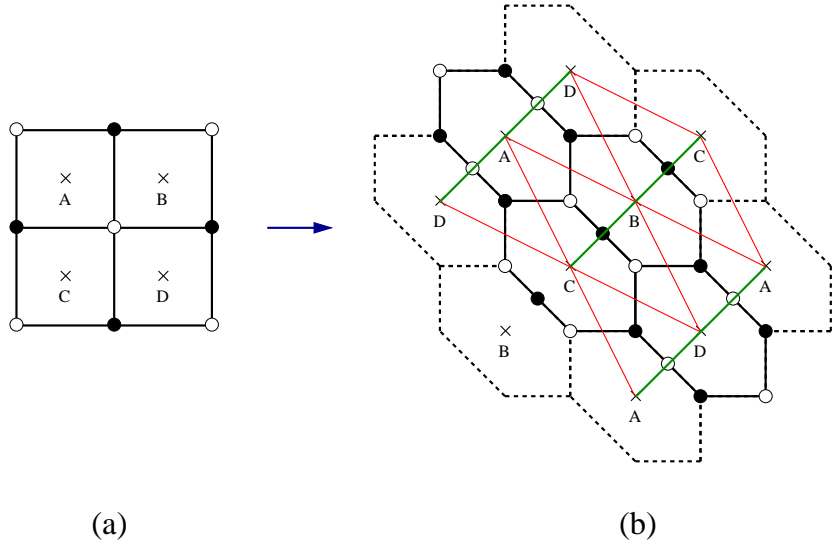}
\caption{The Riemann surface for the $F_0$ is a 2-torus with four punctures. a) For phase I, $\tilde{T}$ contains eight edges and four quartic nodes. b) $\tilde{T}$ and dual triangulation after one possible decomposition of the quartic nodes. Double edges are indicated in green.}
\label{ideal_triangulation_F0_I}
\end{center}
\end{figure}

Let us now revisit $F_0$. In this case, the Riemann surface is a 2-torus with four punctures. We have already discussed the tiling associated to phase II in the previous section. Phase I has eight chiral fields and four quartic superpotential terms. \fref{ideal_triangulation_F0_I} shows its $\tilde{T}$ before \cite{Feng:2005gw} and after decomposing the quartic nodes and the dual ideal triangulation.

\subsection{New Coordinates from Node Decomposition: Experimental Data}

It is interesting to collect explicit examples showing how the rather obvious \eref{dT_gauge_theory} works in practice when decomposing high valence nodes. In the table below, we present several examples showing the agreement between the geometric and field counting (in the decomposed theory) determinations of $3 m$. In the superpotential column, we simply indicate the terms that are not cubic.
\beq
\begin{array}{|c||c|c||c|c|c|}
\hline
& \ \ (g,n) \ \ & \ 3 m \mbox{ geometry} \ & \ \ \mbox{$N_f$} \ \ & W &  \ 3 m \mbox{ field counting} \ \\ \hline
\mathbb{C}^3 & (0,3) & 3 & 3 & - & 3 \\ \hline
\ \ \ \mbox{conifold} \ \ \ & (0,4) & 6 & 4 & \mbox{2 quartic} & 6 \\ \hline
F_0^{(I)} & (1,4) & 12 & 8 & \mbox{4 quartic} & 12 \\ \hline
F_0^{(II)} & (1,4) & 12 & 12 & - & 12 \\ \hline
dP_1 & (1,4) & 12 & 10 & \mbox{2 quartic} & 12 \\ \hline
dP_3^{(I)} & (1,6) & 18 & 12 & \ \ \begin{array}{c} \mbox{3 quartic +} \\ \mbox{1 sextic} \end{array} \ \ & 18 \\ \hline
\end{array}
\nonumber
\eeq
\medskip
Of course, in all cases $m$ is equal to $N_g$ of the gauge theory, too.
\medskip

\subsection*{Infinite families of examples: $L^{a,b,a}$ and $Y^{p,q}$}

We can certainly continue exploring multiple examples. Instead of doing so, we conclude this section considering $L^{a,b,a}$ and $Y^{p,q}$ {\it infinite} families of theories. 

\begin{figure}[h]
\begin{center}
\includegraphics[width=5cm]{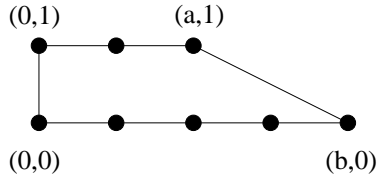}
\caption{Toric diagram for the real cones over $L^{a,b,a}$ manifolds.}
\label{toric_Laba}
\end{center}
\end{figure}

The toric diagram for $L^{a,b,a}$ is shown in \fref{toric_Laba}. They have $g=0$ and $n=a+b+2$ (i.e. they give rise to spheres with an arbitrary number of punctures). We thus have $3m=3(a+b)$.

Let us now focus on the corresponding quiver gauge theories \cite{Franco:2005sm}. They have $(a+b)$ gauge groups. There are two bifundamental fields pointing in opposite directions connecting each consecutive pair of nodes. In addition, $(b-a)$ nodes also have an adjoint field.  We thus have $N_f=a+3b$. The superpotential contains cubic terms and $(2a)$ quartic terms. Plugging this information into \eref{dT_gauge_theory}, we obtain $3m=(a+3b)+2a=3(a+b)$, reproducing the geometric result.

\begin{figure}[h]
\begin{center}
\includegraphics[width=4.5cm]{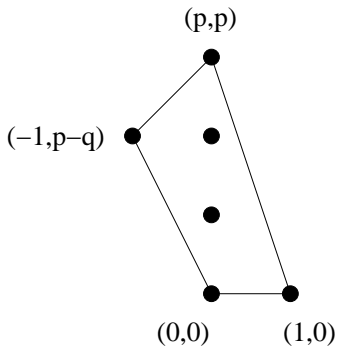}
\caption{Toric diagram for the real cones over $Y^{p,q}$ manifolds.}
\label{toric_Ypq}
\end{center}
\end{figure}

The toric diagram for $Y^{p,q}$ is shown in \fref{toric_Ypq}. They have $g=p-1$ and $n=4$. As a result, we have $3m=6p$. The associated gauge theories have been introduced in \cite{Benvenuti:2004dy}. They can be constructed iteratively starting from $Y^{p,p}$. The cone over $Y^{p,p}$ is the $\mathbb{C}^3/\mathbb{Z}_{2p}$ orbifold. As a result, the corresponding gauge theory has $2p$ gauge groups, $6p$ fields and purely cubic superpotential, trivially satisfying $3m=6p$. The gauge theory for $Y^{p,q}$ corresponds to adding $(p-q)$ impurities. Every impurity removes two fields and introduces two quartic terms. Thus, we conclude that field counting in the decomposed theory correctly reproduces the value of $3m$ for the entire $Y^{p,q}$ family.

\subsection{Kashaev coordinates}

We close this section with a brief digression discussing how another set of coordinates in Teichm\"uller space, Kashaev coordinates \cite{Kashaev}, fit nicely into our construction. Indeed, Kashaev coordinates are naturally defined in terms of the dual of an ideal triangulation, namely $\tilde{T}$. Our discussion will be in terms of 3-valent nodes. This follows from decomposing high valence nodes and including double edges.

\begin{figure}[h]
\begin{center}
\includegraphics[width=3cm]{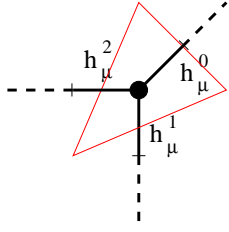}
\caption{Kashaev coordinates corresponding to the three half edges ending on node $\mu$. They are subject to the constraint \eref{constraint_Keshaev}. The associated triangulation is shown in red. A blue asterisk indicates the decorated corner. The opposite half edge in the dual graph can be determined in terms of the other two using the constraint \eref{constraint_Keshaev}.}
\label{Keshaev_node}
\end{center}
\end{figure}

To define these coordinates, the first step is to split every edge in $\tilde{T}$ into two halves. Interestingly, as we pointed out in Section \ref{section_mirror}, the double line implementation of zig-zag paths naturally leads to such decomposition. To each node $\mu$ in $\tilde{T}$ ($\mu=1,\ldots, 2m$), equivalently to each triangle in the ideal triangulation, we associate three real coordinates $h^s_\mu$ ($s=0,1,2$) corresponding to the half edges connected to it as shown in \fref{Keshaev_node}.
In this way we obtain $6m$ coordinates, two for each edge in the triangulation.

Kashaev coordinates obey the following constraint at each node

\beq
\prod_s h^s_\mu = 1 \, .
\label{constraint_Keshaev}
\eeq
This means that, at every node, we can choose one of the three edges and solve it in terms of the other two using \eref{constraint_Keshaev}, resulting in $4m$ coordinates. The standard way of encoding the choice of $2m$ coordinates that are solved for is by means of {\it decorated ideal triangulations}, in which one corner of each ideal triangle is singled out (decorated). The coordinate that is eliminated is the one associated to the half edge in $\tilde{T}$ located opposite to the decorated corner. 

The relation between Fock and Kashaev coordinates becomes intuitive in this picture. For and edge $x_i$ connecting nodes $\mu$ and $\nu$, we have
\beq
x_i=h^s_\mu h^t_\nu \, ,
\eeq
where $h^s_\mu$ and $h^t_\nu$ are the coordinates for the two halves of $x_i$.

It would be extremely interesting to investigate how Kashaev algebra and quantization arise in the context of dimer models. We leave this question for future work and focus on Fock coordinates in the reminder of the paper.

\section{Quantum Teichm\"uller from Dimers}

\label{section_QT_from_dimers}

In this section we explain how (a natural generalization of) the Chekhov-Fock commutation rules for quantizing Teichm\"uler space give rise to Goncharov-Kenyon commutators. 

\subsection{Commutation Relations: Goncharov-Kenyon from Chekhov-Fock}

\begin{figure}[h]
\begin{center}
\includegraphics[width=15cm]{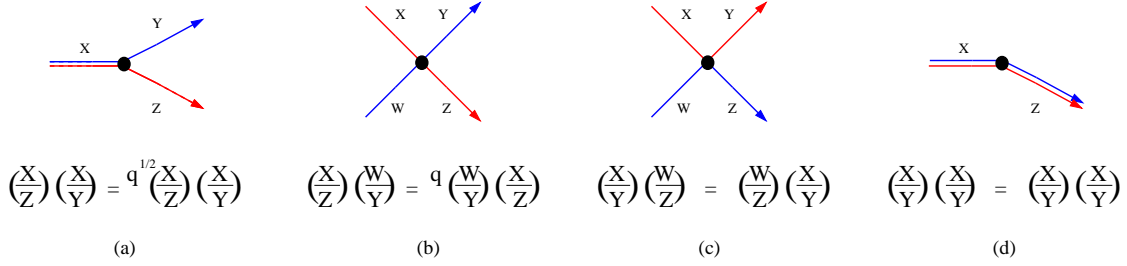}
\caption{Local contributions to Poisson brackets in terms of partial paths. We show $\gamma_1$ and $\gamma_2$ in red and blue, respectively.}
\label{local_rules_commutators_2}
\end{center}
\end{figure}

In order to make the connection, we first re-interpret \fref{local_rules_commutators} in terms of partial paths using the prescription in Section \ref{section_integrals_closed_paths}. The result is shown in \fref{local_rules_commutators_2}.

We will prove the equivalence in two steps. We will first consider the case of purely cubic tilings, which only involves rules (a) and (d) and then proceed to the general case, for which (b) and (c) must be included. Rule (d) is trivially satisfied, so we focus on the other three.

\subsubsection{Purely cubic case}

Goncharov and Kenyon determine that the commutator associated to the cubic vertex in \fref{local_rules_commutators_2}.a is
\beq
\left. \left({X\over Z}\right)\left({X \over Y}\right)\right|_{GK}=q^{1 \over 2} \left({X \over Y}\right)\left({X\over Z}\right)\,.
\label{rule_a_GK}
\eeq

\begin{figure}[h]
\begin{center}
\includegraphics[width=8.5cm]{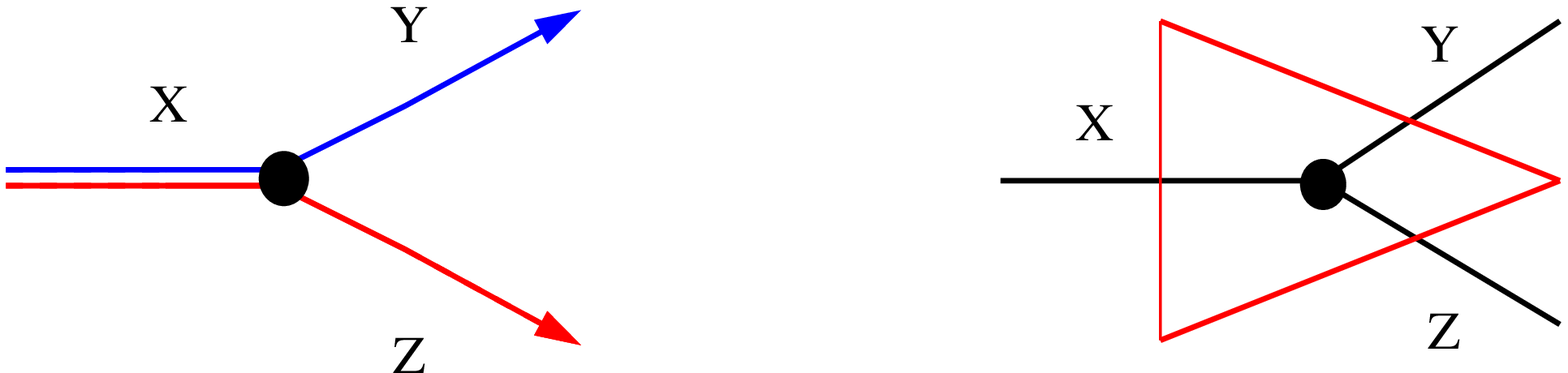}
\caption{The cubic rule (a) from \fref{local_rules_commutators_2} and its triangulation on the Riemann surface.}
\label{triangulation_cubic_node}
\end{center}
\end{figure}

To compute the same commutator using Chekhov-Fock rules, we first dualize the corresponding vertex of $\tilde{T}$ as in \fref{triangulation_cubic_node} and then determine the individual commutators using the prescription in \fref{rhombus_Fock_rules}. We obtain

\beq
\left. \left({X\over Z}\right)\left({X \over Y}\right)\right|_{CF}=q^{(0)_{XX}-(-1)_{XY}-(-1)_{ZX}+(1)_{ZY}}\left({X \over Y}\right)\left({X\over Z}\right)=q^3\left({X \over Y}\right)\left({X\over Z}\right) \, ,
\label{rule_a_Fock}
\eeq
where we have used subindices to indicate the individual contributions to the exponent coming from commuting pairs of variables. We will adhere to this notation in the rest of the paper.

We conclude that CF and GK quantizations agree for purely cubic tilings after a harmless choice of the relative values of $\hbar$ betwee Fock and GK quantizations

\beq
\hbar_{CF} = 6 \, \hbar_{GK}\,.
\label{scaling_F_GK}
\eeq
This relative scaling, which we have proved in full generality at the level of local commutation rules, is exhibited in various explicit examples in Section \eref{section_examples_cubic}.

\subsubsection{The general case}

Matching of the cubic node fix the relative value of $\hbar$ between CF and GK quantization as summarized in the scaling \eref{scaling_F_GK}. Below, we apply this scaling to the (b) and (c) commutators we want to compute.

\subsubsection*{Generalizing Fock rules}

\begin{figure}[h]
\begin{center}
\includegraphics[width=13cm]{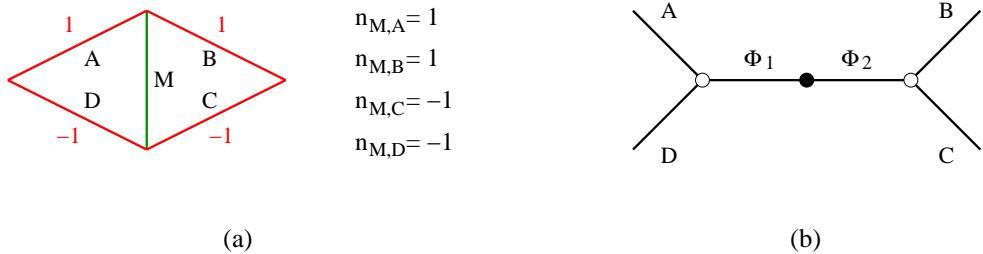}
\caption{a) This rhombus diagram summarizes the contributions to $n_{e,e'}$, which define the generalization of Poisson brackets between coordinates in Teichm\"uller space that deals with double edges. The Poisson brackets are promoted to commutators in the quantum theory.  This generalization is necessary to make contact with GK commutators. b) The choice of signs corresponds to defining $M=\Phi_1/\Phi_2$, with the edges corresponding to $\Phi_1$ and $\Phi_2$ in the $\tilde{T}$ graph dual to the rhombus sitting at the left and right of the figure, respectively.}
\label{rhombus_generalized_Fock_rules}
\end{center}
\end{figure}

Brane tilings with arbitrary k-valent nodes require the addition of rules (b) and (c). We can analyze a quartic node by decomposing it by the insertion of a 2-valent one.\footnote{As mentioned in Section \ref{section_integrable_dimers}, there might be additional edges terminating on the nodes in \fref{local_rules_commutators_2}. These edges are not important for our analysis because they are not part of the two paths under consideration. As done in the previous section, they can be isolated from the quartic node using 2-valent nodes.} In order to make contact with the results of Goncharov and Kenyon, it is necessary to generalize commutation relations by the addition of the rule in \fref{rhombus_generalized_Fock_rules} for double edges. The need for a new rule is a reflection of the fact rules (b) and (c) are genuinely independent of (a) and cannot be derived from it.

Like the distinction of double edges, this new rule is only necessary if one is interested in making contact with dimer models, for which bipartiteness is a crucial property. The mild difference between the new rule and \fref{rhombus_Fock_rules} is indeed quite reasonable. Recall that $M=\Phi_1/\Phi_2$, i.e. the right half of a double edge (which corresponds to $\Phi_2$) enters the definition of $M$ with a negative power. As a result, it is natural to flip the signs of commutators involving the right triangle.

A natural question is whether it is possible to identify double edges given just an ideal triangulation of a Riemann surface. At this time, we do not have a full answer to it. That said, it is straightforward to identify a necessary condition for a triangulation not to have double edges: its dual graph must be bipartite. For example, ideal triangulations consisting of an odd number of triangles can only be connected to dimer models after the addition of double edges.

\subsubsection*{Reproducing Goncharov-Kenyon}

We now use the new rule to reproduce Goncharov-Kenyon rules (b) and (c). It is instructive to go over the details of the calculation using, in each case, the two possible decompositions of the quartic node.

\subsubsection*{Rule b}

\begin{figure}[h]
\begin{center}
\includegraphics[width=12cm]{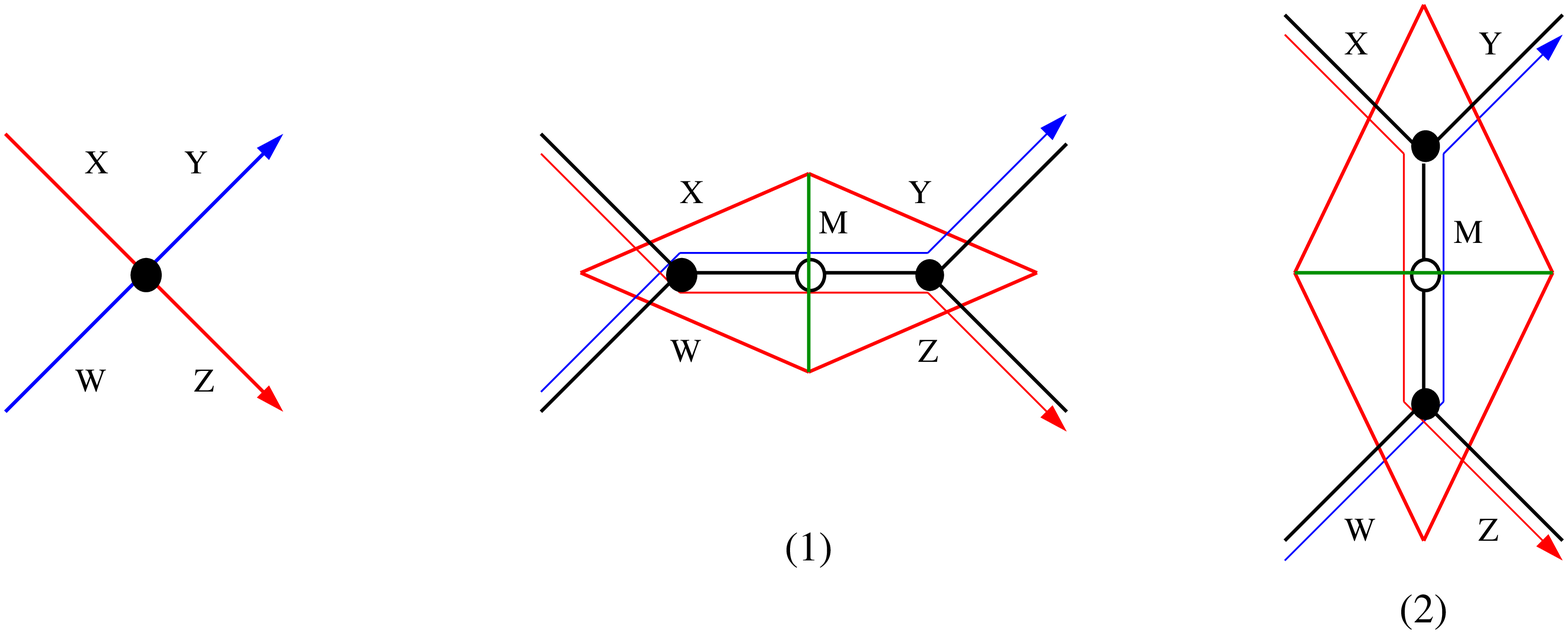}
\caption{Vertex associated with rule (b) in \fref{local_rules_commutators} and its two possible decompositions.}
\label{splitting_quartic_node_1}
\end{center}
\end{figure}

\fref{splitting_quartic_node_1} shows the two possible decompositions of the quartic node with their corresponding triangulations.

Let us consider the first decomposition. Both the $\left({X\over Z}\right)$ and $\left({W \over Y}\right)$ paths go through the double edge in the same direction so they get extended to $\left({X\over ZM}\right)$  and $\left({W \over YM}\right)$. The commutation relation we want to compute becomes

\beq
\hspace{-.35cm}\begin{array}{ccl}
\left. \left({X\over ZM}\right)\left({W \over YM}\right)\right|_{CF}&=&q^{\left[(1)_{XW}-(0)_{XY}-(-1)_{XM}-(0)_{ZW}+(1)_{ZY}+(1)_{ZM}-(-1)_{MW}+(1)_{MY}+(0)_{MM}\right]} \left({W \over YM}\right) \left({X\over ZM}\right) \\ \\
& = & q^6 \left({W \over YM}\right) \left({X\over ZM}\right) \, ,
\end{array}
\eeq
which is the expected result, given the relative normalization in \eref{scaling_F_GK}.

Now consider the second decomposition of the quartic node. Interestingly, the two extended paths traverse the double edge in opposite directions. As a result, $\left({X\over Z}\right)$ and $\left({W \over Y}\right)$ become $\left({X\over ZM}\right)$ and $\left({W M\over Y}\right)$, which results in

\beq
\hspace{-.65cm}\begin{array}{ccl}
\left. \left({X\over ZM}\right)\left({W M\over Y}\right)\right|_{CF}&=&q^{\left[(0)_{XW}-(-1)_{XY}+(1)_{XM}-(-1)_{ZW}+(0)_{ZY}-(-1)_{ZM}-(-1)_{MW}+(1)_{MY}-(0)_{MM}\right]} \left({W M\over Y}\right) \left({X\over ZM}\right)\\ \\
& = & q^6 \left({W M\over Y}\right) \left({X\over ZM}\right) .
\end{array}
\eeq
Once again, we obtain the expected result.

\subsubsection*{Rule c}

\begin{figure}[h]
\begin{center}
\includegraphics[width=12cm]{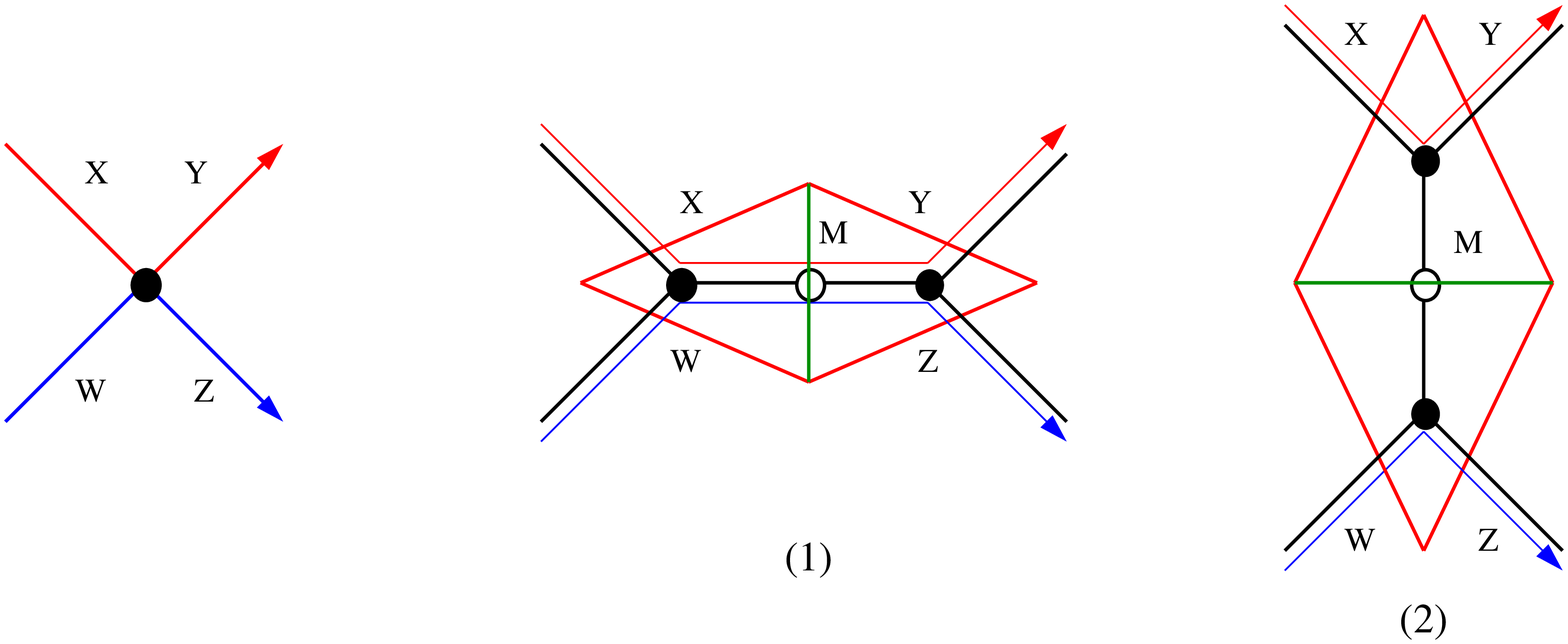}
\caption{Vertex associated with rule (c) in \fref{local_rules_commutators} and its two possible decompositions.}
\label{splitting_quartic_node_2}
\end{center}
\end{figure}

Rule (c) can be similarly proved. \fref{splitting_quartic_node_2} shows the two possible decompositions of the node.

For the first decomposition, $\left({X\over Y}\right)$ and $\left({W \over Z}\right)$ become $\left({X\over YM}\right)$ and $\left({W \over ZM}\right)$ so we have

\beq
\hspace{-.7cm}\begin{array}{ccl}
\left. \left({X\over YM}\right)\left({W \over ZM}\right)\right|_{CF}&=&q^{\left[(1)_{XW}-(0)_{XZ}-(-1)_{XM}-(0)_{YW}+(-1)_{YZ}+(-1)_{YM}-(-1)_{MW}+(-1)_{MZ}+(0)_{MM}\right]}  \left({W \over ZM}\right) \left({X\over YM}\right)  \\ \\
& = & q^0 \left({W \over ZM}\right) \left({X\over YM}\right) .
\end{array}
\eeq

In the second decomposition, the paths do not involve the double edge and we have

\beq
\begin{array}{ccl}
\left. \left({X\over Y}\right)\left({W \over Z}\right)\right|_{CF}&=&q^{\left[(1)_{XW}-(0)_{XZ}-(0)_{YW}+(-1)_{YZ}\right]} \left({W \over Z}\right) \left({X\over Y}\right) \\ \\
& = & q^0 \left({W \over Z}\right) \left({X\over Y}\right)  .
\end{array}
\eeq

\subsection{Triangulation Flips from Seiberg Duality (Urban Renewal)}

\begin{figure}[h]
\begin{center}
\includegraphics[width=10cm]{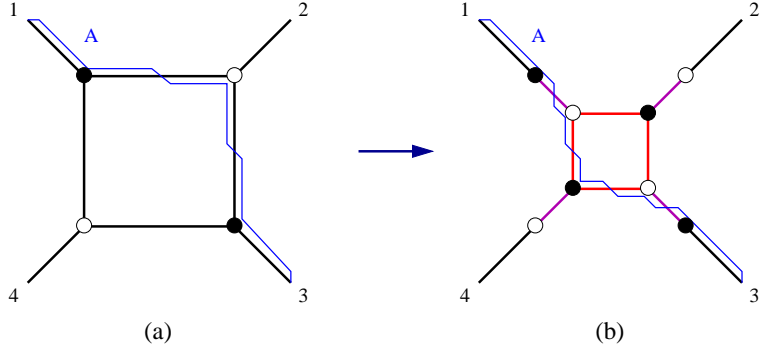}
\caption{Urban renewal transformation (Seiberg duality) of $T$. New edges are shown in color. Purple and red edges correspond to mesons and dual quarks, respectively. A zig-zag path in double line notation is shown in blue.}
\label{urban_renewal}
\end{center}
\end{figure}

Urban renewal is an important transformation of brane tilings. From a quiver perspective, it corresponds to Seiberg duality on gauge groups with an equal number of colors and flavors $\cite{Franco:2005rj}$. Such gauge groups are represented by squares in $T$. \fref{urban_renewal} shows the urban renewal transformation, where we have indicated new edges in color. Purple and red edges correspond to mesons and dual quarks in the Seiberg dual theory, respectively.

The effect of urban renewal on zig-zag paths is most clearly seen using double line notation. \fref{urban_renewal} focuses on one of the four zig-zags going through the square, showing it undergoes a ``reflection". Path A originally goes from leg 1 to 3 passing through the vertex connected to leg 2. After the reflection, it passes through 4 when going from 1 to 3. The other three zig-zag paths experience a similar transformation. 

\begin{figure}[h]
\begin{center}
\includegraphics[width=14cm]{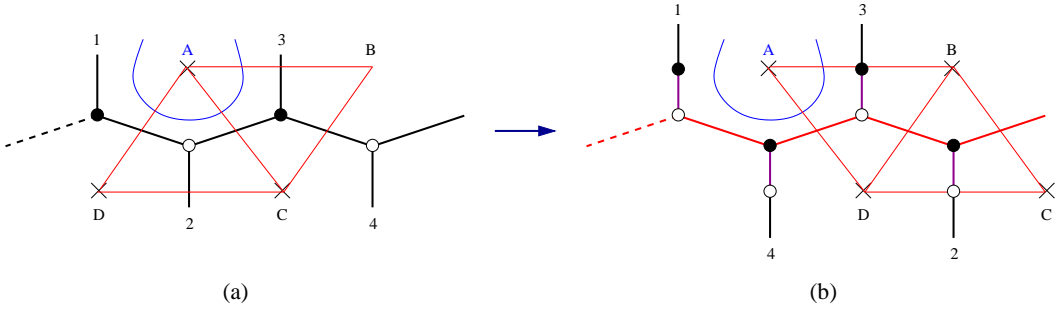}
\caption{A dualized square in $T$ is mapped to a length-4 zig-zag path in $\tilde{T}$. Here we show the effect of the urban renewal transformation on $\tilde{T}$. The horizontal cyclic ordering of the punctures is changed, giving rise to a flip of the triangulation.}
\label{urban_renewal_Ttilde}
\end{center}
\end{figure}

The dualized square face of $T$ is maped to a length-4 zig-zag path in $\tilde{T}$, as shown in \fref{urban_renewal_Ttilde}.a. Zig-zag paths of $T$ map to faces around punctures in $\tilde{T}$. The reflection described in the previous paragraph translates into a shift of legs 1 and 3 with respect to legs 2 and 4. As a result, punctures C and D are also shifted with respect to A and B, changing their horizontal cyclic ordering as shown in \fref{urban_renewal_Ttilde}.b.

Let us consider the rhombus whose vertices, when going clockwise, are given by the sequence $ABCD$. \fref{urban_renewal_Ttilde} shows this rhombus before and after urban renewal. The other rhombi involving the zig-zag path associated to the dualized gauge group transform in the same way. We conclude that urban renewal maps to a triangulation flip in $\Sigma$.

\section{Examples}

\label{section_examples}

In Section \ref{section_QT_from_dimers}, we have introduced a generalization of CF rules and provided a general proof that it implies GK commutators. In this section we present various explicit examples, since we consider it is instructive to see the details of how things work in various specific models. 

\subsection{Purely Cubic Theories}

\label{section_examples_cubic}

\subsubsection{$dP_0$: $(g,n)=(1,3)$}

\begin{figure}[h]
\begin{center}
\includegraphics[width=9cm]{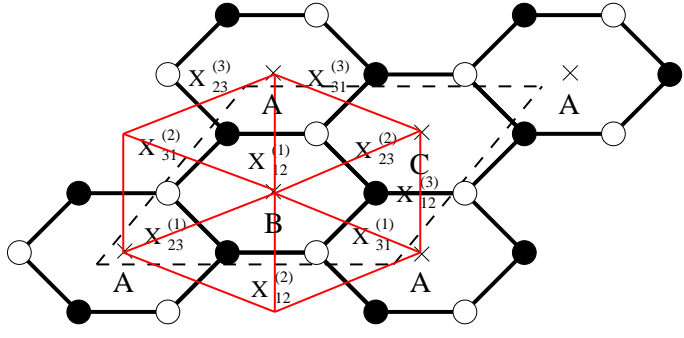}
\caption{$\tilde{T}$ and dual ideal triangulation for $dP_0$.}
\label{Ttilde_dP0}
\end{center}
\end{figure}

\fref{Ttilde_dP0} shows $\tilde{T}$ and the dual triangulation for $dP_0$, which in this case takes the same form of $T$. The gauge theory and $T$ for $dP_0$ can be found in the Appendix. $\Sigma$ is a sphere with three punctures. We have labeled edges in order to compute their commutators. 

Bifundamental fields obey $X_i X_j = q^{n_{\rm{bifund},ij}} X_j X_i$. Applying CF prescription as summarized in \fref{rhombus_Fock_rules}, we determine the matrix $n_{\rm{bifund}}$ to be:

{\footnotesize
\beq
n_{\rm{bifund}}|_{CF}=\left(\begin{array}{c|ccccccccc} 
& \ X_{12}^{(1)} \ & \ X_{12}^{(2)} \ & \ X_{12}^{(3)} \ & \ X_{23}^{(1)} \ & \ X_{23}^{(2)} \ & \ X_{23}^{(3)} \ &  \ X_{31}^{(1)} \ & \ X_{31}^{(2)} \ & \ X_{31}^{(3)} \ \\ \hline
\ X_{12}^{(1)} \ & 0 & 0 & 0 & 0 & 1 & 1 & 0 & -1 & -1  \\
\ X_{12}^{(2)} \ & 0 & 0 & 0 & 1 & 0 & 1 & -1 & 0 & -1 \\
\ X_{12}^{(3)} \ & 0 & 0 & 0 & 1 & 1 & 0 & -1 & -1 & 0 \\
\ X_{23}^{(1)} \ & 0 & -1 & -1 & 0 & 0 & 0 & 0 & 1 & 1 \\
\ X_{23}^{(2)} \ & -1 & 0 & -1 & 0 & 0 & 0 & 1 & 0 & 1 \\
\ X_{23}^{(3)} \ & -1 & -1 & 0 & 0 & 0 & 0 & 1 & 1 & 0 \\
\ X_{31}^{(1)} \ & 0 & 1 & 1 & 0 & -1 & -1 & 0 & 0 & 0 \\
\ X_{31}^{(2)} \ & 1 & 0 & 1 & -1 & 0 & -1 & 0 & 0 & 0 \\
\ X_{31}^{(3)} \ & 1 & 1 & 0 & -1 & -1 & 0 & 0 & 0 & 0 
\end{array}\right)
\label{PB_X_dP0}
\eeq}

Following Section \ref{section_integrals_closed_paths}, the basis of cycles is given by

\beq
\begin{array}{ccccccccccc}
w_1 & = & {X^{(1)}_{12} X^{(2)}_{12} X^{(2)}_{12}\over X^{(1)}_{31} X^{(2)}_{31} X^{(2)}_{31}} & \ \ \ \ & w_2 & = & {X^{(1)}_{23} X^{(2)}_{23} X^{(2)}_{23} \over X^{(1)}_{12} X^{(2)}_{12} X^{(2)}_{12}} & \ \ \ \ & w_3 & = & {X^{(1)}_{31} X^{(2)}_{31} X^{(2)}_{31} \over X^{(1)}_{23} X^{(2)}_{23} X^{(2)}_{23}} \\ \\
z_1 & = & {X_{23}^{(3)}X_{31}^{(3)}\over X_{12}^{(1)}X_{12}^{(2)}} & & 
z_2 & = & {X_{12}^{(1)}X_{12}^{(3)} \over X_{23}^{(2)}X_{31}^{(2)}}
\end{array}
\label{cycles_dP0}
\eeq
Similarly, closed cycles obey $w_i w_j = q^{n_{\rm{cycles},ij}} w_j w_i$. Using \eref{PB_X_dP0}, we obtain

\beq
{1\over 6}\, n_{\rm{cycles}}|_{CF}= \left(\begin{array}{c|ccccc} 
& \ w_1 \ & \ w_2 \ & \ w_3 \ & \ z_1 \ & \ z_2 \\ \hline
\ w_1 \ & 0 & 3 & -3 & 1 & -1 \\
\ w_2 \ & -3 & 0 & 3 & 1 & -1 \\
\ w_3 \ & 3 & -3 & 0 & -2 & 2 \\
\ z_1 \ & -1 & -1 & 2 & 0 & 0 \\
\ z_2 \ & 1 & 1 & -2 & 0 & 0
\end{array}\right)
\label{PB_w_F0_II}
\eeq 
which are precisely the commutators that follow from GK prescription. In the previous equation, we have already introduced the $1/6$ normalization between CF and GK from \eref{scaling_F_GK}.  

\subsubsection{Phase II of $F_0$: $(g,n)=(1,4)$}

\begin{figure}[h]
\begin{center}
\includegraphics[width=4.5cm]{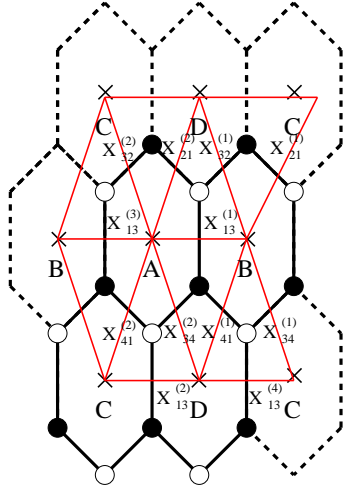}
\caption{$\tilde{T}$ and dual ideal triangulation for phase II of $F_0$.}
\label{Ttilde_F0_II}
\end{center}
\end{figure}

In \fref{Ttilde_F0_II} we show again $\tilde{T}$ and the dual triangulation for phase II of $F_0$,  this time including edge labels. We have considered this theory throughout paper. The corresponding gauge theory and $T$ can be found in the Appendix. $\Sigma$ is a 2-torus with four punctures. 

Using CF rules, we compute:
{\footnotesize
\beq
\hspace{-.5cm} n_{\rm{bifund}}|_{CF}=\left(\begin{array}{c|cccccccccccc} 
& \ X_{13}^{(1)} \ & \ X_{13}^{(2)} \ & \ X_{13}^{(3)} \ & \ X_{13}^{(4)} \ & \ X_{21}^{(1)} \ & \ X_{21}^{(2)} \ &  \ X_{41}^{(1)} \ & \ X_{41}^{(2)} \ & \ X_{32}^{(1)} \ & \ X_{32}^{(2)} \ & \ X_{34}^{(1)} \ & \ X_{34}^{(2)} \ \\ \hline
\ X_{13}^{(1)} \ & 0 & 0 & 0 & 0 & 0 & -1 & -1 & 0 & 1 & 0 & 0 & 1 \\
\ X_{13}^{(2)} \ & 0 & 0 & 0 & 0 & 0 & -1 & 0 & -1 & 0 & 1 & 0 & 1 \\
\ X_{13}^{(3)} \ & 0 & 0 & 0 & 0 & -1 & 0 & 0 & -1 & 0 & 1 & 1 & 0 \\
\ X_{13}^{(4)} \ & 0 & 0 & 0 & 0 & -1 & 0 & -1 & 0 & 1 & 0 & 1 & 0 \\
\ X_{21}^{(1)} \ & 0 & 0 & 1 & 1 & 0 & 0 & 0 & 0 & -1 & -1 & 0 & 0 \\
\ X_{21}^{(2)} \ & 1 & 1 & 0 & 0 & 0 & 0 & 0 & 0 & -1 & -1 & 0 & 0 \\
\ X_{41}^{(1)} \ & 1 & 0 & 0 & 1 & 0 & 0 & 0 & 0 & 0 & 0 & -1 & -1 \\
\ X_{41}^{(2)} \ & 0 & 1 & 1 & 0 & 0 & 0 & 0 & 0 & 0 & 0 & -1 & -1 \\
\ X_{32}^{(1)} \ & -1 & 0 & 0 & -1 & 1 & 1 & 0 & 0 & 0 & 0 & 0 & 0 \\
\ X_{32}^{(2)} \ & 0 & -1 & -1 & 0 & 1 & 1 & 0 & 0 & 0 & 0 & 0 & 0 \\
\ X_{34}^{(1)} \ & 0 & 0 & -1 & -1 & 0 & 0 & 1 & 1 & 0 & 0 & 0 & 0 \\
\ X_{34}^{(2)} \ & -1 & -1 & 0 & 0 & 0 & 0 & 1 & 1 & 0 & 0 & 0 & 0 
\end{array}\right)
\label{PB_X_F0_II}
\eeq}

The basis of cycles if given by
\beq
\begin{array}{ccccccccccc}
w_1 & = & {X_{13}^{(1)} X_{13}^{(2)} X_{13}^{(3)} X_{13}^{(4)}\over X_{41}^{(1)} X_{41}^{(2)} X_{21}^{(1)} X_{21}^{(2)}} & \ \ \ &
w_2 & = & {X_{21}^{(1)} X_{21}^{(2)} \over X_{32}^{(1)} X_{32}^{(2)}} & \ \ \ & z_1 & = & {X_{41}^{(1)} X_{32}^{(1)} \over X_{13}^{(1)} X_{13}^{(4)}} \\ \\ 
w_3 & = & {X_{32}^{(1)} X_{32}^{(2)} X_{34}^{(1)} X_{34}^{(2)} \over X_{13}^{(1)} X_{13}^{(2)} X_{13}^{(3)} X_{13}^{(4)}} & \ \ \ &
w_4 & = & {X_{41}^{(1)} X_{41}^{(2)} \over X_{34}^{(1)} X_{34}^{(2)}} & \ \ \ & z_2 & = & { X_{13}^{(3)} X_{13}^{(4)} \over X_{21}^{(1)} X_{34}^{(1)}}
\end{array}
\label{cycles_F0_II}
\eeq
From \eref{PB_X_F0_II}, we compute
\beq
{1\over 6}\, n_{\rm{cycles}}|_{CF}= \left(\begin{array}{c|cccccc} 
& \ w_1 \ & \ w_2 \ & \ w_3 \ & \ w_4 \ & \ z_1 \ & \ z_2 \ \\ \hline
\ w_1 \ & 0 & -2 & 4 & -2 & 1 & -1\\
\ w_2 \ & 2 & 0 & -2 & 0 & -1 & 1\\
\ w_3 \ & -4 & 2 & 0 & 2 & 1 & -1\\  
\ w_4 \ & 2 & 0 & -2 & 0 & -1 & 1 \\
\ z_1 \ & -1 & 1 & -1 & 1 & 0 & 0 \\
\ z_1 \ & 1 & -1 & 1 & -1 & 0 & 0 
\end{array}\right)
\label{PB_w_F0_II}
\eeq
which is again in perfect agreement with GK.

\bigskip

\subsection{Beyond Cubic Theories}

We now consider examples containing nodes with valence greater than 3. As explained, we need to decompose high valence nodes, paths are extended by including appropriate powers of double edges and \fref{rhombus_generalized_Fock_rules} is necessary for calculating commutators.

\subsubsection{The Conifold: $(g,n)=(0,4)$}

\begin{figure}[h]
\begin{center}
\includegraphics[width=14cm]{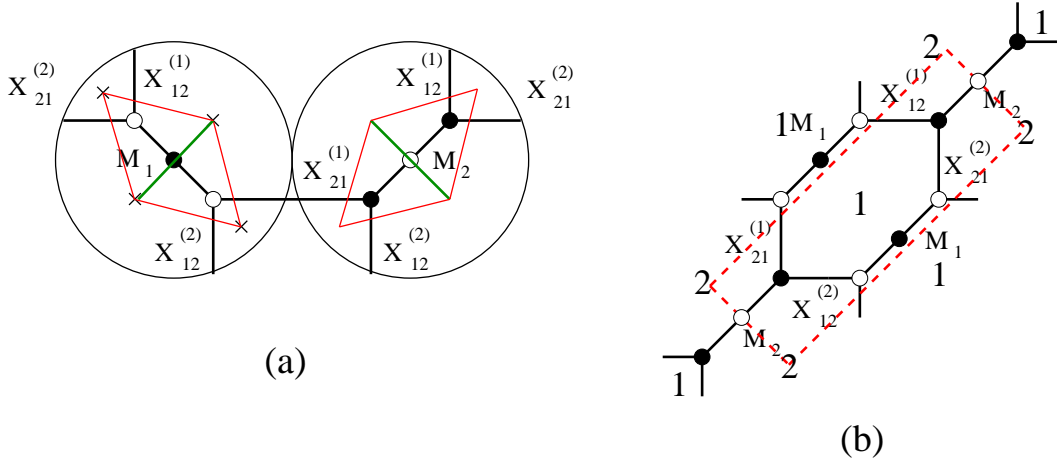}
\caption{The conifold: a) tiling $\tilde{T}$ of $\Sigma$ with double edges and b) effect of the double edges in the original tiling $T$.}
\label{Ttilde_T_conifold}
\end{center}
\end{figure}

\fref{Ttilde_T_conifold}.a shows $\tilde{T}$ for the conifold after one possible decomposition of the two quartic nodes. \fref{Ttilde_T_conifold}.b shows the effect of this decomposition on the original brane tiling.

Using the generalized version of CF rules, we obtain:

{\footnotesize
\beq
n_{\rm{bifund}}|_{CF}= \left(\begin{array}{c|cccccc} 
& \ X_{12}^{(1)} \ & \ X_{12}^{(2)} \ & \ X_{21}^{(1)} \ & \ X_{21}^{(2)} \ & \ M_1 \ & \ M_2 \ \\ \hline
\ X_{12}^{(1)} \ & 0 & 0 & 0 & 0 & -1 & -1 \\
\ X_{12}^{(2)} \ & 0 & 0 & 0 & 0 & 1 & 1 \\
\ X_{21}^{(1)} \ & 0 & 0 & 0 & 0 & -1 & -1 \\
\ X_{21}^{(2)} \ & 0 & 0 & 0 & 0 & 1 & 1 \\
\ M_1 \ & 1 & -1 & 1 & -1 & 0 & 0 \\
\ M_2 \ & 1 & -1 & 1 & -1 & 0 & 0  
\end{array}\right)
\label{PB_X_conifold}
\eeq}

The basic cycles and their extensions after introducing double edges are:
\beq
\begin{array}{ccccccccccc}
w_1 & = & {X^{(1)}_{12} X^{(2)}_{12}\over X^{(1)}_{21} X^{(2)}_{21}} & \ \to \ &  
{X^{(1)}_{12} X^{(2)}_{12} \over X^{(1)}_{21} X^{(2)}_{21}} & \ \ \ \ \ \ & w_2 & = & {X^{(1)}_{21} X^{(2)}_{21} \over X^{(1)}_{12} X^{(2)}_{12}}  & \ \to \ & {X^{(1)}_{21} X^{(2)}_{21} \over X^{(1)}_{12} X^{(2)}_{12}}  \\ \\
z_1 & = & {X^{(1)}_{21} \over X^{(1)}_{12}} & \ \ \ & & & z_2 & = & {X^{(2)}_{21} \over X^{(2)}_{12}}& \ \to \ & {X^{(2)}_{21} M_1 \over X^{(2)}_{12} M_2}
\end{array}
\eeq
Notice that neither $w_1$ nor $w_2$ get powers of $M_1$ or $M_2$ after decomposing the quartic nodes. This is because these paths go through the double edges twice, once in each direction. Then, 
\beq
{1\over 6}\, n_{\rm{cycles}}|_{CF}= \left(\begin{array}{c|cccc} 
& \ w_1 \ & \ w_2 \ & \ z_1 \ & \ z_2 \ \\ \hline
\ w_1 \ & 0 & 0 & 0 & 0 \\
\ w_2 \ & 0 & 0 & 0 & 0 \\
\ z_1 \ & 0 & 0 & 0 & 0 \\  
\ z_2 \ & 0 & 0 & 0 & 0
\end{array}\right)
\label{PB_w_conifold}
\eeq
in agreement with GK. This theory is not too exciting since it is completely non-chiral, which is the reason behind the vanishing of the $\rm{cycles}$ matrix.

\subsubsection{Phase I of $F_0$: $(g,n)=(1,4)$}

\fref{Ttilde_T_conifold}.a shows $\tilde{T}$ for the phase I of $F_0$ after one possible decomposition of the four quartic nodes. \fref{Ttilde_T_conifold}.b shows how the original brane tiling is modified by the decomposition.

\begin{figure}[h]
\begin{center}
\includegraphics[width=13cm]{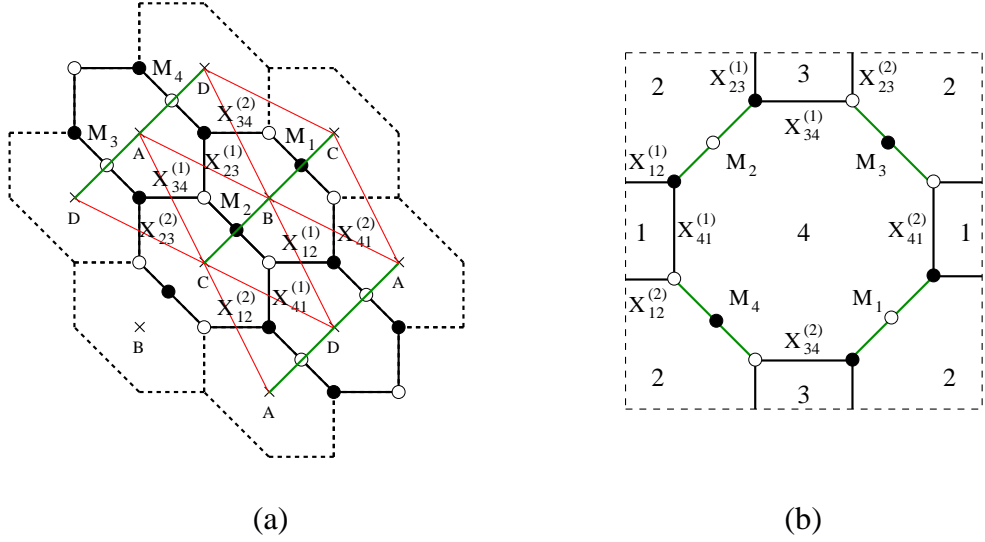}
\caption{Phase I of $F_0$: a) tiling $\tilde{T}$ of $\Sigma$ with double edges and b) effect of the double edges in the original tiling $T$.}
\label{Ttilde_T_F0_I}
\end{center}
\end{figure}

Using the generalized CF rules we compute:
{\footnotesize
\beq
\hspace{-.4cm} n_{\rm{bifund}}|_{CF} = \left(\begin{array}{c|cccccccccccc} 
& \ X_{12}^{(1)} \ & \ X_{12}^{(2)} \ & \ X_{23}^{(1)} \ & \ X_{23}^{(2)} \ & \ X_{34}^{(1)} \ & \ X_{34}^{(2)} \ &  \ X_{41}^{(1)} \ & \ X_{41}^{(2)} \ & \ M_1 \ & \ M_2 \ & \ M_3 \ & \ M_4 \ \\ \hline
\ X_{12}^{(1)} \ & 0 & 0 & 0 & 0 & 0 & 0 & -1 & -1 & 0 & -1 & 1 & 0 \\ 
\ X_{12}^{(2)} \ & 0 & 0 & 0 & 0 & 0 & 0 & -1 & -1 & -1 & 0 & 0 & 1 \\
\ X_{23}^{(1)} \ & 0 & 0 & 0 & 0 & 1 & 1 & 0 & 0 & 0 & -1 & 0 & 1 \\
\ X_{23}^{(2)} \ & 0 & 0 & 0 & 0 & 1 & 1 & 0 & 0 & -1 & 0 & 1 & 0 \\
\ X_{34}^{(1)} \ & 0 & 0 & -1 & -1 & 0 & 0 & 0 & 0 & 0 & 1 & -1 & 0 \\
\ X_{34}^{(2)} \ & 0 & 0 & -1 & -1 & 0 & 0 & 0 & 0 & 1 & 0 & 0 & -1 \\
\ X_{41}^{(1)} \ & 1 & 1 & 0 & 0 & 0 & 0 & 0 & 0 & 0 & 1 & 0 & -1 \\
\ X_{41}^{(2)} \ & 1 & 1 & 0 & 0 & 0 & 0 & 0 & 0 & 1 & 0 & -1 & 0 \\
\ M_1 \ & 0 & 1 & 0 & 1 & 0 & -1 & 0 & -1 & 0 & 0 & 0 & 0 \\
\ M_2 \ & 1 & 0 & 1 & 0 & -1 & 0 & -1 & 0 & 0 & 0 & 0 & 0 \\
\ M_2 \ & -1 & 0 & 0 & -1 & 1 & 0 & 0 & 1 & 0 & 0 & 0 & 0 \\
\ M_2 \ & 0 & -1 & -1 & 0 & 0 & 1 & 1 & 0 & 0 & 0 & 0 & 0
\end{array}\right)
\label{PB_X_F0_I}
\eeq}

The basis of cycles and their extensions by double edges read
\beq
\begin{array}{ccccccccccccccc}
w_1 & = & {X^{(1)}_{12} X^{(2)}_{12}\over X^{(1)}_{41} X^{(2)}_{41}} & \ \ \ \ \ \ & w_2 & = & {X^{(1)}_{23} X^{(2)}_{23}\over X^{(1)}_{12} X^{(2)}_{12}} & \ \to \ & {X^{(1)}_{23} X^{(2)}_{23} M_3 M_4 \over X^{(1)}_{12} X^{(2)}_{12} M_1 M_2} 
& \ \ \ \ \ \ & z_1 & = & {X^{(2)}_{34} \over X^{(2)}_{12}} & \ \to \ & {X^{(2)}_{34} M_4 \over X^{(2)}_{12} M_1} 
\\ \\
w_3 & = & {X^{(1)}_{34} X^{(2)}_{34}\over X^{(1)}_{23} X^{(2)}_{23}} & \ \ \ & w_4 & = & {X^{(1)}_{41} X^{(2)}_{41}\over X^{(1)}_{34} X^{(2)}_{34}} & \ \to \ & {X^{(1)}_{41} X^{(2)}_{41} M_1 M_2 \over X^{(1)}_{34} X^{(2)}_{34} M_3 M_4} 
& \ \ \ \ \ \ & z_2 & = & {X^{(1)}_{41} \over X^{(1)}_{23}} & \ \to \ & {X^{(1)}_{41} M_2 \over X^{(1)}_{23} M_4} 
\end{array}
\eeq

Finally, we compute
\beq
{1\over 6}\, n_{\rm{cycles}}|_{CF}= \left(\begin{array}{c|cccccc} 
& \ w_1 \ & \ w_2 \ & \ w_3 \ & \ w_4 \ & \ z_1 \ & \ z_2 \ \\ \hline
\ w_1 \ & 0 & 2 & 0 & -2 & 1 & -1 \\
\ w_2 \ & -2 & 0 & 2 & 0 & 1 & 1 \\
\ w_3 \ & 0 & -2 & 0 & 2 & -1 & 1 \\  
\ w_4 \ & 2 & 0 & -2 & 0 & -1 & -1 \\
\ z_1 \ & -1 & -1 & 1 & 1 & 0 & 1 \\
\ z_2 \ & 1 & -1 & -1 & 1 & -1 & 0
\end{array}\right)
\label{PB_w_F0_I}
\eeq
Not surprisingly, this also matches GK result.

\section{Conclusions}

We have introduced a correspondence between dimer models and the Teichm\"uller space of Riemann surfaces. We explained how arbitrary dimer models give rise to ideal triangulations via the decomposition of nodes with valence greater than 3. There is a one-to-one correspondence between Fock coordinates and single and double edges in the decomposed tiling, which correspond to single fields and pairs of fields in the quiver, respectively. We showed that the commutators between loops introduced by Goncharov and Kenyon in the context of integrable system can be derived from a natural generalization (necessary to deal with double edges) of Chekhov-Fock rules. Finally, we explained how urban renewal on the brane tiling is mapped to triangulation flips.

There is a web of connections between the models we have studied and other theories that deserves to be studied in detail. First, the integrable systems defined by dimer models also arise from 5d, $\mathcal{N}=1$ gauge theories compactified on a circle \cite{Nekrasov:1996cz,Mina_et_al}. Such gauge theories can be constructed by wrapping M5-branes on $\Sigma$. Sending the circle radius to zero leads to 4d, $\mathcal{N}=2$ gauge theories and is mapped to the non-relativistic limit of the corresponding integrable systems. In this limit, $\Sigma$ becomes the Seiberg-Witten curve of the gauge theory. 

In addition, it would be interesting to investigate the connection between quantum Teichm\"uller and quantum Liouville (see e.g. \cite{Verlinde:1989ua,Teschner:2003at,Teschner:2010je} and the recent work \cite{Terashima:2011qi}) from a dimer model perspective. To do so, we need a dimer understanding of geodesics and length operators. The dimer model techniques introduced in \cite{GarciaEtxebarria:2006aq} for studying arbitrary resolutions of toric singularities seem to be well suited for describing pants decomposition of Riemann surfaces and might be relevant for this purpose. Establishing a link between our story and Liouville theory on Riemann surfaces will open yet another connection to certain 4d $\mathcal{N}=2$ gauge theories via the Alday-Gaiotto-Tachikawa proposal of an equivalence between the instanton partition function of the gauge theory on $\mathbb{R}^4$/$S^4$ and conformal blocks/correlation functions of Liouville theory \cite{Alday:2009aq}. It is certainly worth studying all these connections.

It would be interesting to study whether the interpretation in terms of dimer models sheds some new light on the connection between Kashaev and Chekhov-Fock quantizations of Teichm\"uller space.

We plan to investigate these questions in future work.

\bigskip

\section*{Acknowledgments}

We thank M. Aganagic, R. Eager and K. Schaeffer for collaboration on related projects. We thank G. Giribet for useful clarifications on the AGT correspondence. We are particularly thankful to A. Goncharov and R. Kenyon for reading a draft of this paper and sharing \cite{GK} prior to its publication. S. F. is supported by the National Science Foundation under Grant No. PHY05-51164.

\bigskip

\appendix

\section{Gauge Theories and Brane Tilings}

In order to provide a self-contained presentation, we summarize in Table \ref{examples_considered} the gauge theories and brane tilings for the explicit examples considered in the paper. 

\begin{table}
\vspace{-1.3cm}
\hspace{-.6cm}
$
\ba{|c|}
\hline \hline
\mathbb{C}^3 \\ \hline
\ba{c|c|c}
{\epsfxsize=1in\epsfbox{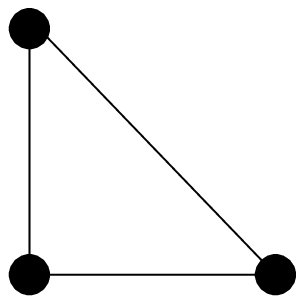}} 
		& 
{\epsfxsize=3.3in\epsfbox{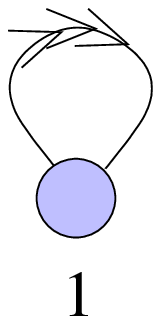}} 
		& {\epsfxsize=2in\epsfbox{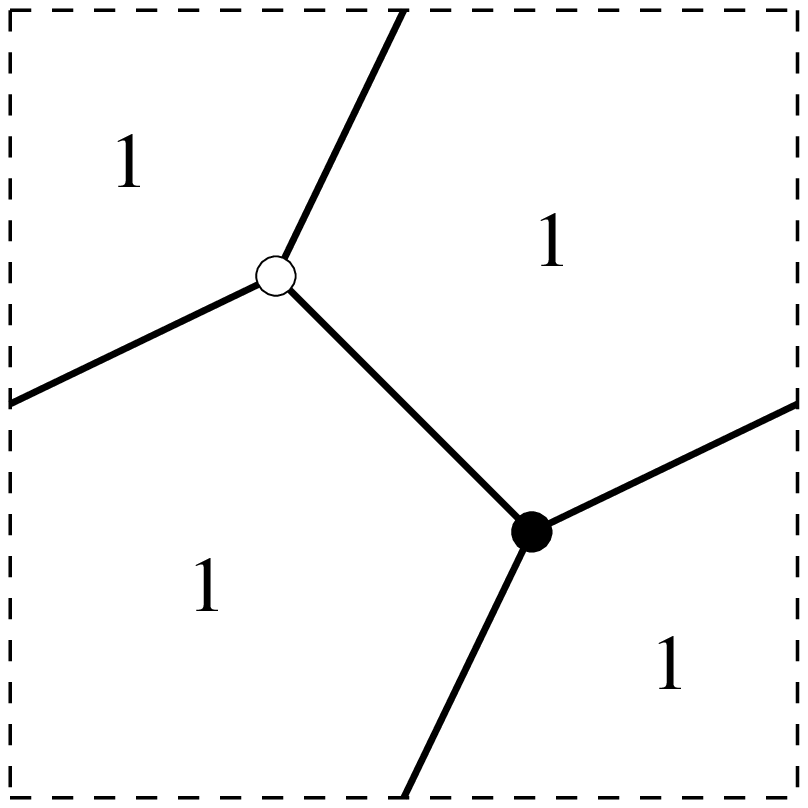}}  \\
& W=\epsilon_{ijk}X^{(i)}X^{(j)}X^{(k)} & 
\ea \\ \hline \hline
\mbox{Conifold}\\ \hline
\ba{c|c|c}
{\epsfxsize=1in\epsfbox{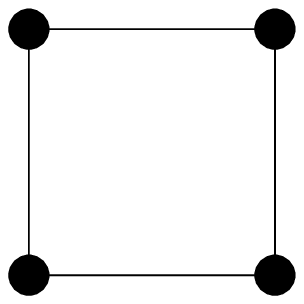}}
		& 
{\epsfxsize=3.3in\epsfbox{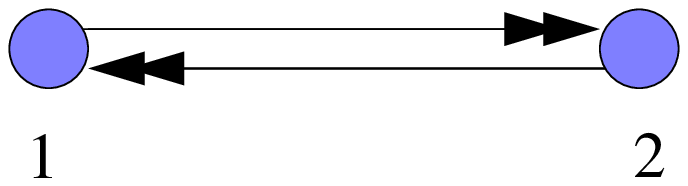}} 
		& {\epsfxsize=2in\epsfbox{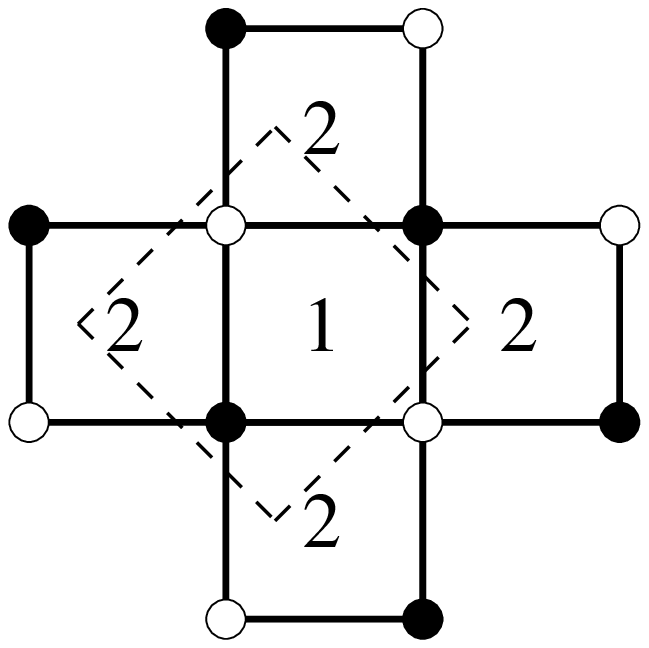}}  \\
& W=\epsilon_{ij}\epsilon_{mn} X_{12}^{(i)}X_{21}^{(m)}X_{12}^{(j)}X_{21}^{(n)} & 
\ea \\ \hline \hline
dP_0 \\ \hline
\ba{c|c|c}
{\epsfxsize=1in\epsfbox{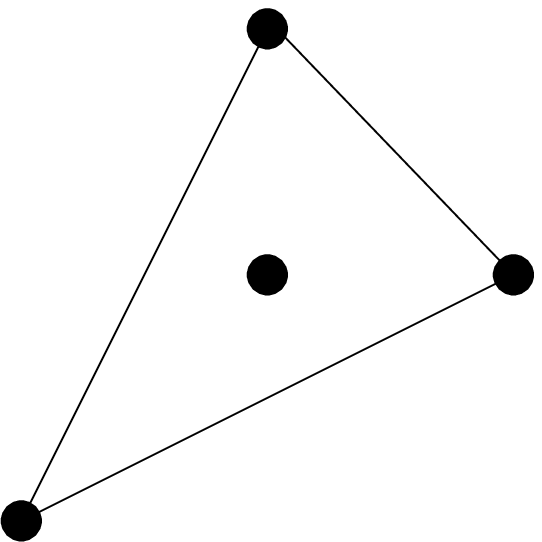}}
		& 
{\epsfxsize=3.3in\epsfbox{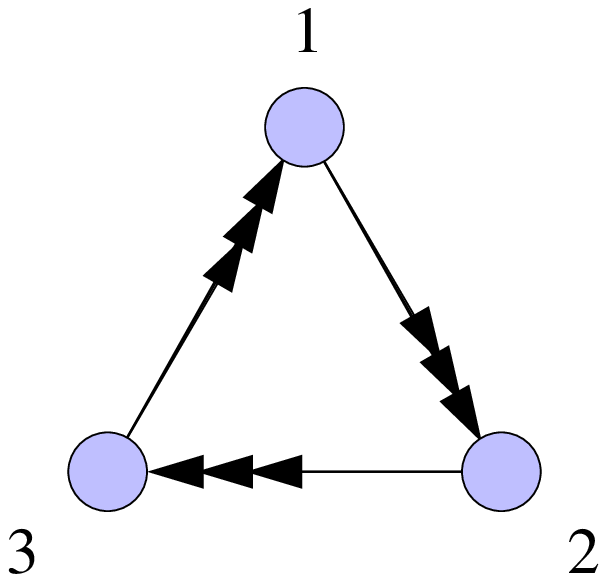}} 
		& {\epsfxsize=2in\epsfbox{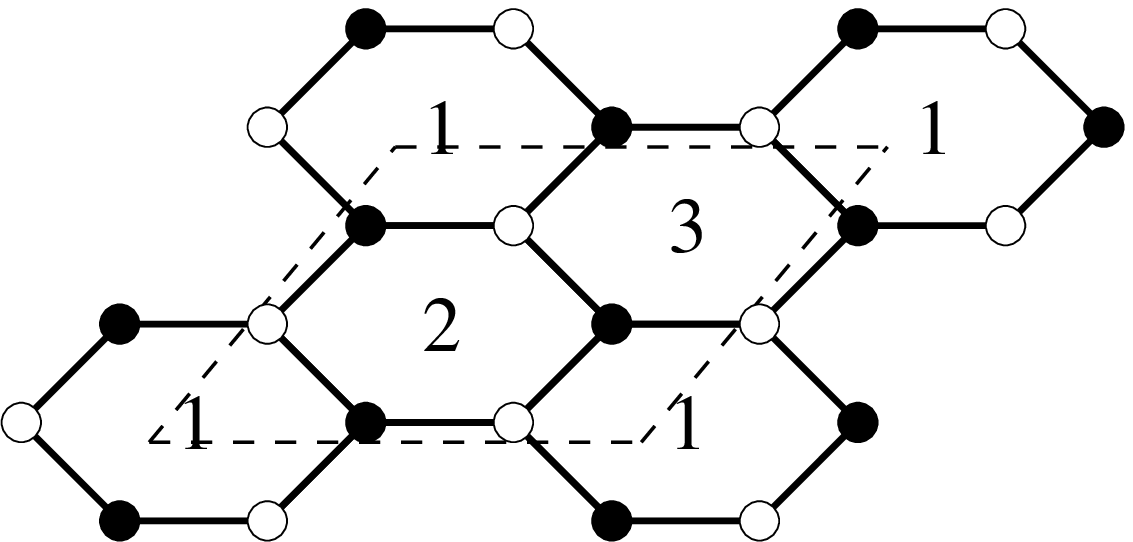}}  \\
& W=\epsilon_{ijk}X_{12}^{(i)}X_{23}^{(j)}X_{31}^{(k)} & 
\ea \\ \hline \hline
F_0 \\ \hline
\mbox{Phase I} \\ \hline
\ba{c|c|c}
{\epsfxsize=1in\epsfbox{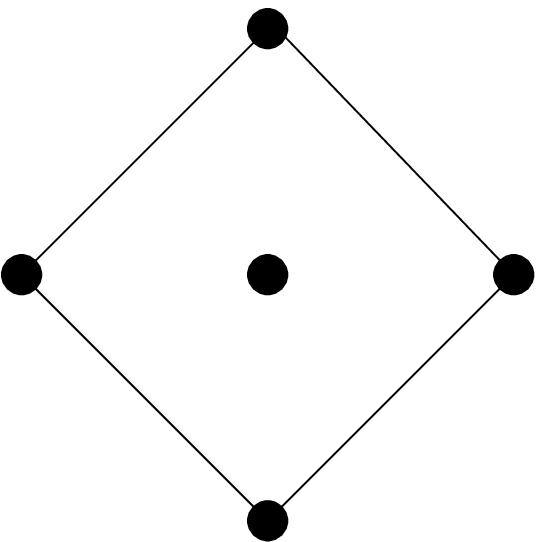}}
		& 
{\epsfxsize=3.3in\epsfbox{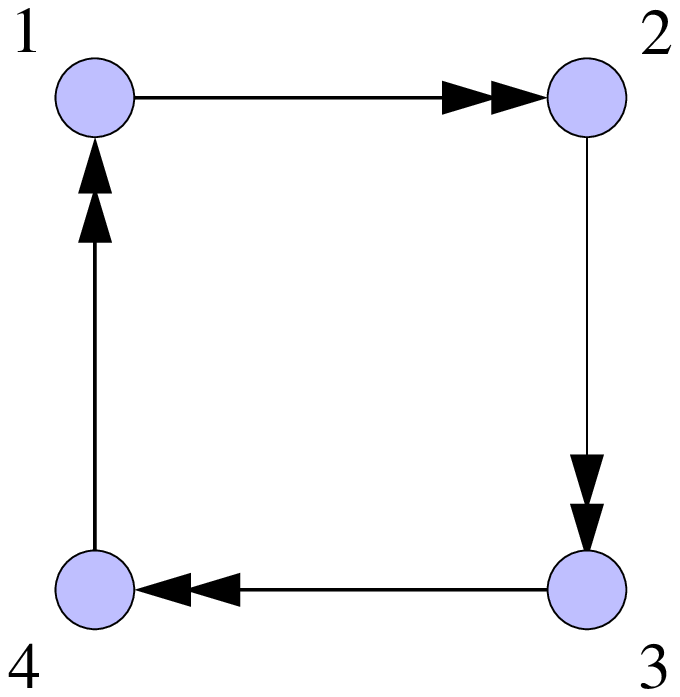}} 
		& {\epsfxsize=2in\epsfbox{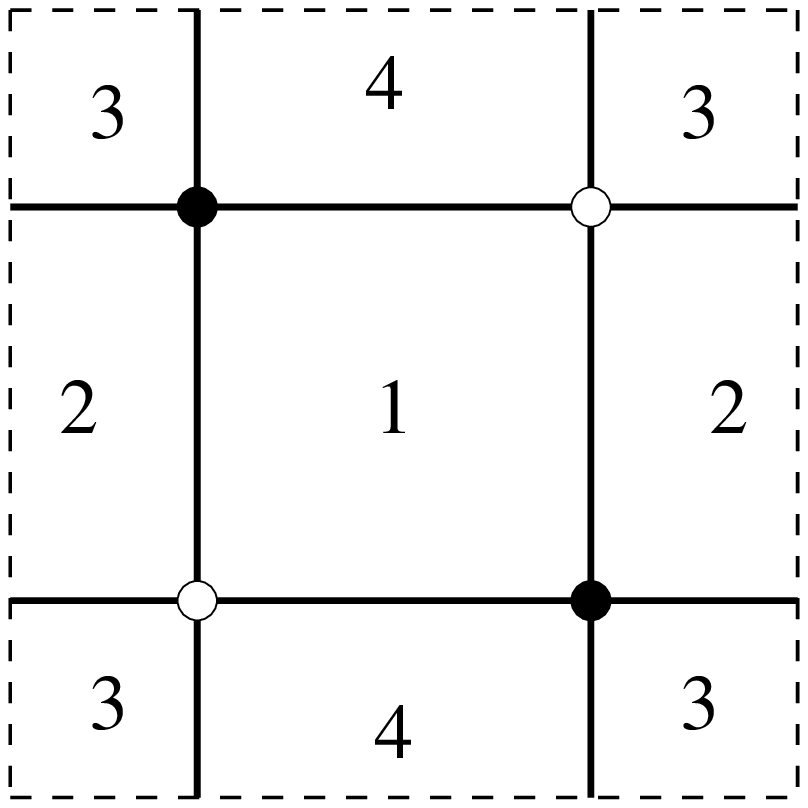}}  \\
& W=\epsilon_{ij}\epsilon_{mn} (X_{13}^{(i,m)}X_{32}^{(j)}X_{21}^{(n)}-X_{13}^{(i,m)}X_{32}^{(j)}X_{21}^{(n)}) & 
\ea \\ \hline
\mbox{Phase II} \\ \hline
\ba{c|c|c}
{\epsfxsize=1in\epsfbox{blank.eps}} 	& 
{\epsfxsize=3.3in\epsfbox{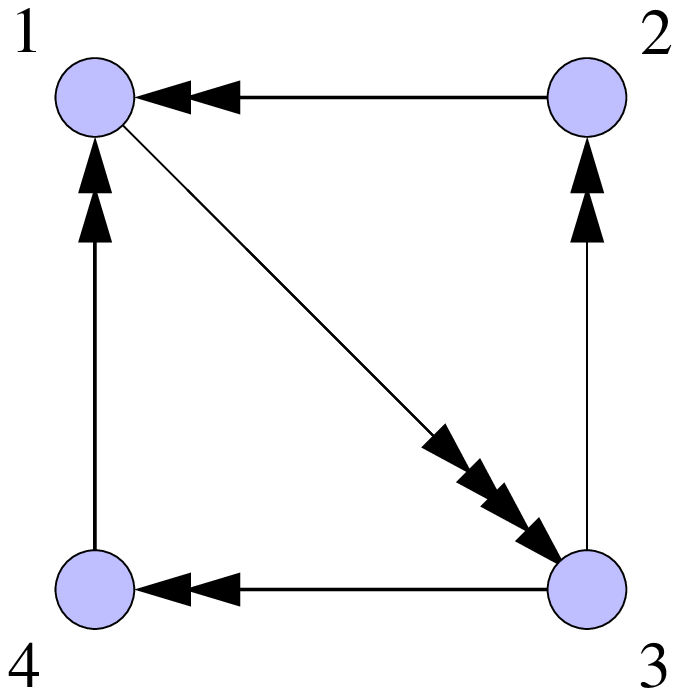}} 
		& {\epsfxsize=2in\epsfbox{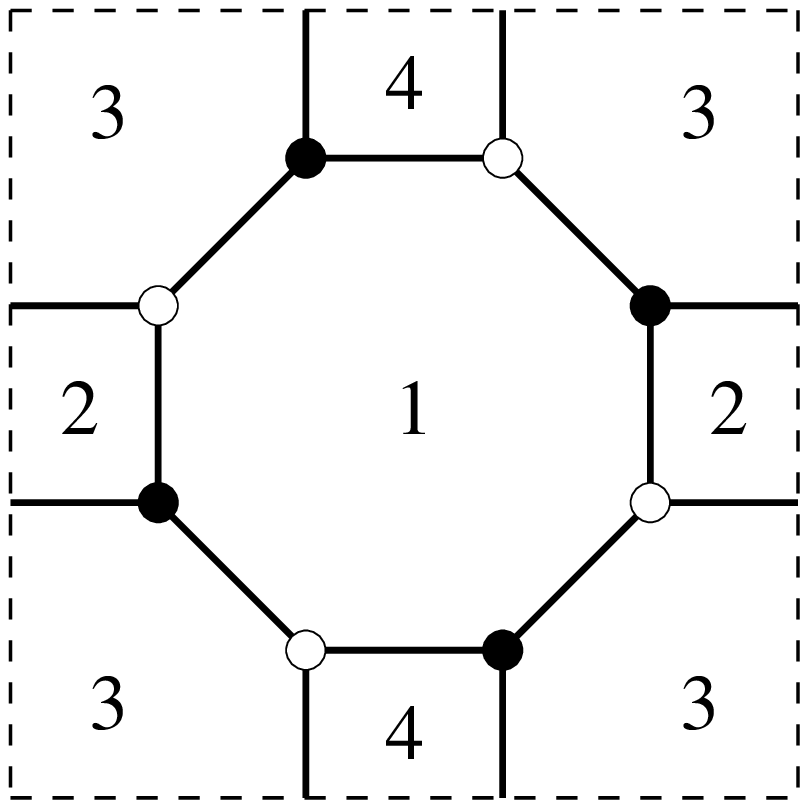}}  \\
& W=\epsilon_{ij}\epsilon_{mn} X_{12}^{(i)}X_{23}^{(m)}X_{34}^{(j)}X_{41}^{(n)} & 
\ea \\ \hline \hline

\ea
$
\caption{\label{examples_considered}Toric diagrams, gauge theories and brane tilings for the explicit examples considered in the paper.}
\end{table}


\end{document}